\documentclass[twocolumn,showpacs]{revtex4}

\usepackage{graphics}

\begin{document}

\title{Nonclassical-light generation in a photonic band-gap nonlinear planar
waveguide}

\author{Jan Pe\v{r}ina, Jr.}
\affiliation{Joint Laboratory of Optics
of Palack\'{y} University and
Institute of Physics of Academy of Sciences of the Czech
Republic, 17. listopadu 50A, 772 07 Olomouc, Czech Republic}
\altaffiliation[]{Faculty of Natural Sciences,
Palack\'{y} University, 17. listopadu 50,
772 07 Olomouc, Czech Republic}
\email{perina_j@sloup.upol.cz}
\author{Concita Sibilia}
\author{Daniela Tricca}
\author{Mario Bertolotti}
\affiliation{INFM-CNR at Dipartimento di Energetica, Universit\`{a} ``La
Sapienza'' di Roma, Via A. Scarpa 16, 00161 Roma, Italy}

\begin{abstract}
Optical parametric process occurring in a photonic band-gap planar
waveguide is studied from the point of view of nonclassical-light
generation. Nonlinearly interacting optical fields are
described by the generalized superposition of coherent signals and
noise using the method of operator linear corrections to a
classical strong solution. Scattered backward-propagating
fields are taken into account. Squeezed light as well as light with
sub-Poissonian statistics can be obtained in two-mode fields under
the specified conditions.
\end{abstract}


\maketitle

\section{Introduction}

During the last years an increasing attention has been devoted
to properties of photonic band-gap structures
\cite{Bertolotti2001,Joannopoulos}. It has been shown that intense
nonlinear processes can occur in such structures built up from
nonlinear materials. For example, second harmonic and sub-harmonic
generation has been predicted and also observed
\cite{Scalora1997,Dumeige2001}. Advantages of these structures
from the point of view of nonlinear interactions are based on high
densities of local optical modes, on spatial localization of
optical modes in confined regions of the structure and in
convenient fulfilling of phase-matching conditions of a given
nonlinear process. Although quantum nonlinear optics in photonic crystals
is promising, investigation of these structures has been
performed mostly in classical domain.  Such structures are
usually very short (typically tens of $ \mu $m) and so they can be
conveniently used in microoptoelectronics as sources of light
in near future. These nonlinear structures can also serve as
sources of light with nonclassical properties, as predicted in
\cite{Sakoda2002}. As discussed in
\cite{Tricca2004a},
second-harmonic-generation process in a planar nonlinear
waveguide with a corrugation on the top of the
waveguide can be used to control squeezing of the fundamental field; the
corrugation reproduces a photonic-band gap
structure. Periodicity of the grating
was selected to give rise to a longitudinal confinement of the
pump field, phase matching of the process was achieved
introducing a spatial modulation of nonlinear susceptibility.

The aim of the paper is to investigate nonclassical properties of
light (squeezing of vacuum fluctuations and sub-Poissonian
photon-number statistics) generated in a nonlinear planar
photonic-band-gap waveguide fabricated in such a way
that a strong optical parametric process
occurs inside. The corrugation of the waveguide is suitably
selected in order to confine longitudinally  signal and idler modes, while a
weak
longitudinal confinement of the pump field is assumed.
Phase matching for the parametric process is achieved due to
the presence of the grating and due to modal dispersion of the
guided wave geometry (see, e.g., in \cite{Cristiani1999}), although
our considerations are not specific to a particular material.

The paper is organized as follows.
Equations describing nonlinear interaction of optical fields
are derived in Sec. 2. Possibility of
squeezed-light generation is discussed in Sec. 3.
Sec. 4 is devoted to generation of light with sub-Poissonian
photon-number statistics. Sec. 5 provides conclusions.
Quantum derivation
of the corresponding equations is contained in Appendix A, whereas
Appendix B deals with commutation relations among operators of
the outgoing fields.

\section{Derivation of equations describing the nonlinear
interaction}

An optical field present inside a planar waveguide can be decomposed
into modes. Then we have for the electric-field amplitude
$ {\bf E}({\bf r},t) $ at point $ {\bf r}=(x,y,z) $ and time $ t $
the following expression:
\begin{eqnarray}    
{\bf E}(x,y,z,t) = i \sum_{m}
\sqrt{\frac{\hbar\omega_m}{2\epsilon_0 \bar{\epsilon}_r V}}
{\bf e}_m \nonumber \\
\times \left[ A_m(z) f_m(x,y) \exp[i({\bf k}_m)_z z -
i\omega_m t] - {\rm h.c.} \right] ,
\label{1}
\end{eqnarray}
where $ A_m $ is amplitude of the $ m $-th mode, $ f_m $ means
transverse eigenfunction of the $ m $-th mode
($ \Delta_{x,y} f_m(x,y) = 0 $), $ {\bf e}_m $
stands for polarization vector, $ \omega_m $ denotes frequency
and $ {\bf k}_m $ is wavevector of the $ m $-th mode.
Mean permittivity of the waveguide is denoted as
$ \epsilon_r $, $ \epsilon_0 $ stands for vacuum
permittivity, $ \hbar $ is reduced Planck constant and $ V $
volume of the structure. The symbol $ {\rm h.c.} $ stands for
hermitian-conjugated terms.

The electric-field amplitude $ {\bf E}({\bf r},t) $ fulfils
the wave equation inside the waveguide:
\begin{equation}     
\nabla^2 {\bf E} - \mu\epsilon_0 \epsilon_r \frac{\partial^2
{\bf E} }{\partial t^2} = \mu \frac{\partial^2
{\bf P}_{\rm nl} }{\partial t^2} ;
\label{2}
\end{equation}
where $ \mu $ denotes vacuum permeability and
$ {\bf P}_{\rm nl} $ describes nonlinear polarization of the
medium. Relative permittivity $ \epsilon_r({\bf r}) $ can
be written as follows:
\begin{equation}     
 \epsilon_r(x,y,z) = \bar{\epsilon}_r(x,y) \left[ 1 +
 \Delta\epsilon_r(x,y,z) \right] .
\end{equation}
Small variations of permeability described by $ \Delta
\epsilon_r(x,y,z) $ are related to the corrugation caused by
a linear grating fabricated on the surface of the waveguide.
It is useful to decompose permittivity variations
$  \Delta\epsilon_r $ as a function of spatial coordinate
$ z $ into harmonic functions:
\begin{equation}    
  \Delta\epsilon_r(x,y,z) = \sum_{q=-\infty}^{\infty}
  \varepsilon_q \exp\left[ iq \frac{2\pi}{\Lambda_l} z \right] ,
 \varepsilon_0 = 0 ;
\end{equation}
$ \varepsilon_q $ are coefficients of the decomposition.
Polarization $ {\bf P}_{\rm nl} $ of the medium is determined
using second-order
susceptibility tensor $ \underline{\bf \chi} $:
\begin{equation}    
 {\bf P}_{\rm nl} = \epsilon_0 \underline{\bf \chi} \cdot {\bf E}
 \cdot {\bf E} ;
 \label{5}
\end{equation}
the symbol $ \cdot $ stands for tensorial multiplication.

Substitution of the expression for $ {\bf E}({\bf r},t) $
in Eq. (\ref{1}) into the wave equation in Eq. (\ref{2})
and the assumption $ \left| \frac{\partial^2 A_m}{\partial z^2}
\right| \ll \left| ({\bf k}_m)_z \frac{\partial A_m}{\partial z}
\right| $ (analog of slowly-varying envelope approximation in
time domain to spatial evolution) results in the system of nonlinear
differential equations for amplitudes $ A_m $ of modes participating in
the nonlinear interaction. We have six modes: signal
forward-propagating mode (with amplitude $ A_{s_F} $), signal
backward-propagating mode ($ A_{s_B} $), idler
forward-propagating mode ($ A_{i_F} $), idler
backward-propagating mode ($ A_{i_B} $), pump
forward-propagating mode ($ A_{p_F} $), and finally pump
backward-propagating mode ($ A_{p_B} $). The system of differential
equations is written as follows:
\begin{eqnarray}   
\frac{dA_{s_F}}{dz} &=& iK_{s}\exp(-i\delta_{s}z) A_{s_B}
 \nonumber \\
 & & \mbox{} + 2K_{F} \exp(i\delta_{F}z) A_{p_F} A^*_{i_F}
\nonumber\\
\frac{dA_{i_F}}{dz} &=& iK_{i}\exp(-i\delta_{i}z) A_{i_B}
 \nonumber \\
 & & \mbox{} + 2K_{F} \exp(i\delta_{F}z) A_{p_F} A^*_{s_F}
\nonumber\\
\frac{dA_{s_B}}{dz} &=& -iK^*_{s}\exp(i\delta_{s}z) A_{s_F}
 \nonumber \\
 & & \mbox{} - 2K_{B} \exp(-i\delta_{B}z) A_{p_B} A^*_{i_B}
 \nonumber\\
\frac{dA_{i_B}}{dz} &=& -iK^*_{i}\exp(i\delta_{i}z) A_{i_F}
 \nonumber \\
 & & \mbox{} - 2K_{B} \exp(-i\delta_{B}z) A_{p_B} A^*_{s_B}
\nonumber\\
 \frac{dA_{p_F}}{dz} &=& -2K^*_{F} \exp(-i\delta_{F}z)
   A_{s_F} A_{i_F}
\nonumber\\
 \frac{dA_{p_B}}{dz} &=& 2K^*_{B} \exp(i\delta_{B}z) A_{s_B}
    A_{i_B}
\label{6}
\end{eqnarray}
and
\begin{eqnarray}    
 \delta_{a} &=& |({\bf k}_{a_F})_z| + |({\bf k}_{a_B})_z| -
 \delta_l ,
  \hspace{0.5cm} a=s,i, \nonumber \\
 \delta_l &=& \frac{2\pi}{\Lambda_l}, \nonumber \\
 \delta_{b} &=& |({\bf k}_{p_b})_z| - |({\bf k}_{s_b})_z| -
 |({\bf k}_{i_b})_z| , \hspace{0.5cm} b=F,B .
\end{eqnarray}
The linear coupling constants $ K_{s} $ and $ K_{i} $ are
given as:
\begin{eqnarray}  
 K_{a} &=& \frac{|({\bf k}_{a_F})_z|}{2} \int dx dy \,
 \varepsilon_1(x,y) f^*_{a_F}(x,y) f_{a_B}(x,y) ,
  \nonumber \\
 & & \hspace{3cm} a=s,i .
\end{eqnarray}
The expressions for the nonlinear coupling constants
$ K_{F} $ and $ K_{B} $ are:
\begin{eqnarray}   
 K_{b} &=& \sqrt{\frac{\mu \hbar \omega_p \omega_s \omega_i}{
 8\bar{\epsilon}^2_r V} }
 \int dx dy \, \underline{\chi} \cdot {\bf e}_p \cdot {\bf e}_s
  \cdot {\bf e}_i \nonumber \\
  & & \mbox{} \times f_{p_b}(x,y) f^*_{s_b}(x,y) f^*_{i_b}(x,y),
  \hspace{0.5cm} b=F,B .
\end{eqnarray}

The same equations as in Eq. (\ref{6}) can be derived using
quantum description of optical fields. The quantum version of Eq.
(\ref{6}) uses operators of electric-field amplitudes
instead of classical electric-field amplitudes (for details,
see Appendix A).
A general solution of the quantum variant of the nonlinear system
of operator equations in Eq. (\ref{6}) can only be found using
a completely numerical approach. However, taking into account
conditions of the nonlinear interaction in a real photonic-band-gap
planar waveguide, we
can apply the approximation of small linear operator corrections to
a classical strong solution for mean values. We can write in this case:
\begin{eqnarray}      
  \hat{A}_a = A_a + \delta \hat{A}_a , \nonumber \\
   a = s_F, i_F, p_F, s_B, i_B, p_B.
\end{eqnarray}
Mean amplitudes $ A_a $ obey classical equations given in Eq. (\ref{6}),
whereas the evolution of small operator corrections $ \delta\hat{A}_a $
is governed by the following equations:
\begin{eqnarray}     
\frac{d\delta \hat{A}_{s_F}}{dz} &=& {\cal K}_{s} \delta\hat{A}_{s_B}
 + {\cal K}_{F}\left[ A_{p_F} \delta \hat{A}^\dagger_{i_F}
 + A^*_{i_F} \delta \hat{A}_{p_F} \right] ,
\nonumber\\
\frac{d\delta\hat{A}_{i_F}}{dz} &=& {\cal K}_{i} \delta\hat{A}_{i_B}
  + {\cal K}_{F} \left[ A_{p_F} \delta\hat{A}^\dagger_{s_F}
 +  A^*_{s_F} \delta\hat{A}_{p_F}  \right] ,
\nonumber\\
\frac{d\delta\hat{A}_{s_B}}{dz} &=& {\cal K}^*_{s} \delta\hat{A}_{s_F}
 - {\cal K}_{B} \left[ A_{p_B} \delta\hat{A}^*_{i_B}
  + A^*_{i_B} \delta\hat{A}_{p_B} \right],
 \nonumber\\
\frac{d\delta\hat{A}_{i_B}}{dz} &=& {\cal K}^*_{i} \delta\hat{A}_{i_F}
 - {\cal K}_{B} \left[ A_{p_B}\delta\hat{A}^\dagger_{s_B}
  + A^*_{s_B}\delta\hat{A}_{p_B} \right],
\nonumber\\
 \frac{d\delta\hat{A}_{p_F}}{dz} &=& -{\cal K}^*_{F}
  \left[ A_{s_F} \delta\hat{A}_{i_F} + A_{i_F}
   \delta\hat{A}_{s_F} \right],
\nonumber\\
 \frac{d\delta\hat{A}_{p_B}}{dz} &=& {\cal K}^*_{B} \left[ A_{s_B}
    \delta\hat{A}_{i_B} + A_{i_B} \delta\hat{A}_{s_B} \right] .
\label{11}
\end{eqnarray}
The constants $ {\cal K}_s $, $ {\cal K}_i $, $ {\cal K}_F $, and
$ {\cal K}_B $ introduced in Eq. (\ref{11}) are defined as:
\begin{eqnarray}   
 {\cal K}_a &=& iK_a \exp(-i\delta_a z). \hspace{1cm} a=s,i,
 \nonumber \\
 {\cal K}_F &=& 2K_F \exp(i\delta_F z),
 \nonumber \\
 {\cal K}_B &=& 2K_B \exp(-i\delta_B z).
\end{eqnarray}

Assuming linear interaction being much stronger than the
nonlinear one ($ K_F A $ and $ K_B A $ are much lower than
$ K_s $ and $ K_i $; $ A $ means an arbitrary classical
amplitude), Eq. (\ref{6}) can be solved analytically. We then
have for the signal and idler modes:
\begin{eqnarray}    
A_{aF}(z) &=& \exp\left( -i\frac{\delta_a z}{2} \right) \nonumber\\
 & &\mbox{} \times
 \left[ A_{a_F}(0) \cos(\Delta_a z) + {\cal C}_a \sin(\Delta_a z)
  \right], \nonumber \\
A_{aB}(z) &=& \exp\left(i\frac{\delta_a z}{2} \right) \nonumber \\
 & & \mbox{} \hspace{-1cm} \times
 \left[ A_{a_F}(0) \left( -\frac{\delta_a}{2K_a} \cos(\Delta_a z)
  + \frac{i\Delta_a}{K_a}\sin(\Delta_a z) \right) \right. \nonumber \\
  & & \mbox{} \hspace{-0.5cm} \left. +
  {\cal C}_a \left( -\frac{\delta_a}{2K_a} \sin(\Delta_a z)
  - \frac{i\Delta_a}{K_a}\cos(\Delta_a z) \right)
  \right], \nonumber \\
   \Delta_a &=& \sqrt{\frac{\delta^2_a}{4}-|K_a|^2},
  \hspace{1cm} a=s,i.
\end{eqnarray}
The symbol $ A_{s_F}(0) $ [$ A_{i_F}(0) $] stands for an incident
signal- [idler-] field classical amplitude. Values of the constants
$ {\cal C}_s $ and
$ {\cal C}_i $ are determined from the conditions $ A_{s_B}(L) = 0 $ and
$ A_{i_B}(L) = 0 $; $ L $ denotes the length of the waveguide.

Classical amplitudes $ A_{p_F} $ and $ A_{p_B} $ of the pump modes
inside the waveguide are given as:
\begin{eqnarray}    
 A_{p_b}(z) &=& A_{p_b}(0) \nonumber \\
  & & \mbox{} \hspace{-1cm} \pm \frac{2K^*_b {\cal E}_s
  {\cal E}_i}{i(\Delta_b-\Delta_s-\Delta_i)} \left(
  \exp[-i(\Delta_b-\Delta_s-\Delta_i)z ] -1 \right)
  \nonumber \\
    & & \mbox{} \hspace{-1cm} \pm \frac{2K^*_b {\cal E}_s
  {\cal F}_i}{i(\Delta_b-\Delta_s+\Delta_i)} \left(
  \exp[-i(\Delta_b-\Delta_s+\Delta_i)z ] -1 \right)
  \nonumber \\
  & & \mbox{} \hspace{-1cm} \pm \frac{2K^*_b {\cal F}_s
  {\cal E}_i}{i(\Delta_b+\Delta_s-\Delta_i)} \left(
  \exp[-i(\Delta_b+\Delta_s-\Delta_i)z ] -1 \right)
  \nonumber \\
   & & \mbox{} \hspace{-1cm} \pm \frac{2K^*_b {\cal F}_s
  {\cal F}_i}{i(\Delta_b+\Delta_s+\Delta_i)} \left(
  \exp[-i(\Delta_b+\Delta_s+\Delta_i)z ] -1 \right) ,
  \nonumber \\
  & & \hspace{3cm} b=F,B .
\label{14}
\end{eqnarray}
The upper signs in Eq. (\ref{14}) are appropriate for $ A_{p_F} $
whereas the lower signs stand for $ A_{p_B} $.
The constants occurring in Eq. (\ref{14}) are given as:
\begin{eqnarray}    
 \Delta_F &=& \delta_F + \frac{\delta_s}{2} +
  \frac{\delta_i}{2}  , \nonumber \\
 \Delta_B &=& - \delta_B - \frac{\delta_s}{2} -
  \frac{\delta_i}{2} \nonumber \\
  {\cal E}_{a_F} &=& \frac{1}{2} \left( A_{a_F}(0)-i{\cal C}_a
    \right) \nonumber \\
  {\cal F}_{a_F} &=& \frac{1}{2} \left( A_{a_F}(0)+i{\cal C}_a
    \right),  \nonumber \\
  {\cal E}_{a_B} &=& \frac{2\Delta_a-\delta_a}{4K_a}
     \left(A_{a_F}(0) -i {\cal C}_a \right) \nonumber \\
  {\cal F}_{a_B} &=& \frac{2\Delta_a+\delta_a}{4K_a}
   \left(-A_{a_F}(0) -i {\cal C}_a \right), \hspace{0.5cm} a=s,i.
\end{eqnarray}
The constant $ A_{p_F}(0) $ characterizes the incident pump-field
amplitude and value of the constant $ A_{p_B}(0) $ is determined
from the condition $ A_{p_B}(L)=0 $.

Solution of the system of linear equations in Eq. (\ref{11}) for operator
corrections can be found numerically and expressed in the matrix form:
\begin{equation}     
 \pmatrix{\delta \hat{\cal A}_{F,\rm out} \cr \delta \hat{\cal
 A}_{B,\rm in}}
 =  \pmatrix{ {\cal U}_{FF} & {\cal U}_{FB} \cr
 {\cal U}_{BF} & {\cal U}_{BB} }
 \pmatrix{\delta \hat{\cal A}_{F,\rm in} \cr
  \delta \hat{\cal A}_{B,\rm out}},
 \label{16}
\end{equation}
where
\begin{eqnarray}      
\delta \hat{\cal A}_{F,\rm in} = \pmatrix{\delta \hat{A}_{s_F}(0) \cr
\delta \hat{A}^\dagger_{s_F}(0) \cr \delta \hat{A}_{i_F}(0) \cr
\delta \hat{A}^\dagger_{i_F}(0) \cr \delta \hat{A}_{p_F}(0) \cr
\delta \hat{A}^\dagger_{p_F}(0)} , \hspace{0.5cm}
\delta \hat{\cal A}_{F,\rm out} = \pmatrix{ \delta \hat{A}_{s_F}(L) \cr
\delta \hat{A}^\dagger_{s_F}(L) \cr \delta \hat{A}_{i_F}(L) \cr
\delta \hat{A}^\dagger_{i_F}(L) \cr \delta \hat{A}_{p_F}(L) \cr
\delta \hat{A}^\dagger_{p_F}(L)} , \nonumber \\
\delta \hat{\cal A}_{B,\rm in} = \pmatrix{ \delta \hat{A}_{s_B}(L) \cr
\delta \hat{A}^\dagger_{s_B}(L) \cr \delta \hat{A}_{i_B}(L) \cr
\delta \hat{A}^\dagger_{i_B}(L) \cr \delta \hat{A}_{p_B}(L) \cr
\delta \hat{A}^\dagger_{p_B}(L) }, \hspace{0.5cm}
\delta \hat{\cal A}_{B,\rm out} = \pmatrix{ \delta \hat{A}_{s_B}(0) \cr
\delta \hat{A}^\dagger_{s_B}(0) \cr \delta \hat{A}_{i_B}(0) \cr
\delta \hat{A}^\dagger_{i_B}(0) \cr \delta \hat{A}_{p_B}(0) \cr
\delta \hat{A}^\dagger_{p_B}(0) } .
\label{17}
\end{eqnarray}
The matrices $ {\cal U}_{FF} $, $ {\cal U}_{FB} $,
$ {\cal U}_{BF} $, and $ {\cal U}_{BB} $ are determined by
numerical solution of Eq. (\ref{11}). Precision of numerical
solution of Eq. (\ref{11}) can be monitored using identities
stemming from commutation relations among operators (see Appendix B).

Input-output relations among linear operator corrections can be found
solving Eq. (\ref{16}) with respect to vectors $ \delta
\hat{\cal A}_{F,\rm out} $ and $ \delta \hat{\cal A}_{B,\rm out} $:
\begin{eqnarray}     
 \pmatrix{\delta \hat{A}_{F,\rm out} \cr
 \delta \hat{A}_{B,\rm out}}  &=&
 \pmatrix{ {\cal U}_{FF}-{\cal U}_{FB} {\cal U}^{-1}_{BB}
 {\cal U}_{BF} & {\cal U}_{FB} {\cal U}^{-1}_{BB} \cr
 -{\cal U}^{-1}_{BB}{\cal U}_{BF} & {\cal U}^{-1}_{BB}}
 \nonumber \\
 & & \mbox{} \times
 \pmatrix{\delta \hat{\cal A}_{F,\rm in} \cr \delta \hat{\cal
 A}_{B,\rm in}}.
\end{eqnarray}
The output linear operator corrections contained in vectors
$ \delta \hat{\cal A}_{F,\rm out} $ and $ \delta \hat{\cal
A}_{B,\rm out} $
obey bosonic commutation relations provided that the
input linear operator corrections in vectors
$ \delta \hat{\cal A}_{F,\rm in} $
and $ \delta \hat{\cal A}_{B,\rm in} $ obey bosonic commutation
relations. It has been shown in \cite{Luis1996} that this
nontrivial property is fulfilled by any system described by
a quadratic hamiltonian.

Eq. (\ref{11}) can be solved also iteratively assuming
weak linear ($ {\cal K}_s L \ll 1 $,
$ {\cal K}_i L \ll 1 $) as well as nonlinear
($ {\cal K}_F A_{p_F} L \ll 1 $, $ {\cal K}_B A_{p_B} L \ll 1 $)
interactions among modes in the planar waveguide
(weak-interaction approximation).
The obtained expressions provide a useful information about
the behaviour of physical quantities of interest (see the next
sections).

\section{Squeezed-light generation}

Squeezing of fluctuations of an optical field below the level
characterizing vacuum fluctuations can be indicated by mean values of
the variances of quadrature components $ \hat{q}_j $
[$ \hat{q}_j = \hat{A}_j + \hat{A}^\dagger_j $, $ \hat{A}_j $
stands for electric-field-amplitude operator of mode $ j $] and
$ \hat{p}_j $ [$ \hat{p}_j = -i(\hat{A}_j - \hat{A}^\dagger_j) $]
in mode $ j $. Maximum amount of available squeezing measured under
a suitably chosen value of the local-oscillator phase in
homodyne-measurement scheme is given by principal squeeze variance
$ \lambda_j $ \cite{Luks1988}.

The above-introduced quantities can be
generalized to optical fields composed of two modes; we
have $ \hat{q}_{jk} = \hat{q}_j + \hat{q}_k $ and
$ \hat{p}_{jk} = \hat{p}_j + \hat{p}_k $ for quadrature
components of the field composed of modes $ j $ and $ k $.
An optical field described by a compound mode can
be obtained in an output port of a beamsplitter that combines
two single modes from its inputs.

We assume that the interacting fields can be described in
the framework of the generalized superposition of signal and noise
\cite{Perina1991}
(coherent states, squeezed states as well as noise can be considered)
and then we have \cite{PerinaJr2000}:
\begin{eqnarray}     
\langle (\Delta \hat{q}_j)^2\rangle &=& 1+ 2[B_j+{\rm Re}(C_j)],
 \nonumber \\
\langle (\Delta \hat{p}_j)^2\rangle &=& 1+ 2[B_j-{\rm Re}(C_j)],
 \nonumber \\
\lambda_j &=& 1+ 2[B_j-|C_j|], \\
\langle (\Delta \hat{q}_{jk})^2\rangle &=& 2\left[1+ B_j+B_k -
  2{\rm Re}(\bar{D}_{jk}) \right. \nonumber \\
  & & \mbox{} \left. + {\rm Re}( C_j+C_k+2D_{jk}) \right],
  \nonumber \\
\langle (\Delta \hat{p}_{jk})^2\rangle &=& 2\left[1+ B_j+B_k -
  2{\rm Re}(\bar{D}_{jk}) \right. \nonumber \\
  & & \mbox{} \left. - {\rm Re}( C_j+C_k+2D_{jk}) \right],
  \nonumber \\
\lambda_{jk} &=& 2\left[1+ B_j+B_k -
  2{\rm Re}(\bar{D}_{jk}) \right. \nonumber \\
  & & \mbox{} \left. - | C_j+C_k+2D_{jk}| \right] .
\end{eqnarray}
Symbol $ \langle \;\; \rangle $ denotes quantum statistical mean
value.
Quantities $ B_j $, $ C_j $, $ D_{jk} $, and $ \bar{D}_{jk} $
are defined as \cite{PerinaJr2000}:
\begin{eqnarray}   
B_j &=& \langle \Delta\hat{A}^\dagger_j \Delta\hat{A}_j \rangle ,
  \nonumber \\
C_j &=& \langle (\Delta\hat{A}_j)^2 \rangle ,
  \nonumber \\
D_{jk} &=& \langle \Delta\hat{A}_j \Delta\hat{A}_k \rangle ,
 \hspace{1cm} j \neq k,  \nonumber \\
\bar{D}_{jk} &=& -\langle \Delta\hat{A}^\dagger_j \Delta\hat{A}_k,
\rangle \hspace{1cm} j \neq k;
\label{21}
\end{eqnarray}
$ \Delta\hat{A}_j = \hat{A}_j - \langle\hat{A}_j\rangle $.
Quantities defined in Eq. (\ref{21}) can be expressed in terms of
the matrices $ u $ and $ v $ defined in Appendix B and using
incident values of $ B_{j,\rm in,{\cal A}} $ and $ C_{j,\rm in,{\cal A}} $
related to antinormal ordering of field operators (for details,
see \cite{PerinaJr2000}):
\begin{eqnarray}   
 B_{j,\rm in,{\cal A}} &=& \cosh^2(r_{j}) + n_{ch,j} ,
  \nonumber \\
 C_{j,\rm in,{\cal A}} &=& \frac{1}{2} \exp (i\vartheta_j)
  \sinh (2r_j) .
\end{eqnarray}
Symbol $ r_j $ denotes squeeze parameter of the incident $ j $-th
mode, $ \vartheta_j $ means squeeze phase, and $ n_{ch,j} $
stands for mean number of incident chaotic photons.
Values of $ \langle (\Delta \hat{q}_j)^2\rangle $,
$ \langle (\Delta \hat{p}_j)^2\rangle $, and $ \lambda_j $
less than one mean squeezing in single-mode case.
Squeezed light is generated in a compound-mode (two-mode) case
if values of $ \langle (\Delta \hat{q}_{jk})^2\rangle $,
$ \langle (\Delta \hat{p}_{jk})^2\rangle $, or $ \lambda_{jk} $
are less than two.

When discussing properties of the nonlinearly interacting modes
we use symmetry based on the exchange of signal and idler
modes. Quantities and properties that can be derived from this
symmetry are not mentioned explicitly.

Discussion of squeezing is based on the investigation of
principal squeeze variances because they give the maximally
allowed amount of squeezing reachable in an experiment.

\subsection{Weak-interaction approximation}

Assuming incident coherent states (or vacuum
states) principal squeeze variances of single modes
have the following form in weak-interaction approximation:
\begin{eqnarray}     
 \lambda_{s_F} &=& 1 + 2 | I_{p_F} |^2 , \nonumber \\
 \lambda_{s_B} &=& 1 + 2 | I_{p_B} |^2 , \nonumber \\
 \lambda_{p_F} &=& 1 , \nonumber \\
 \lambda_{p_B} &=& 1 ,
\label{22}
\end{eqnarray}
and
\begin{eqnarray}      
I_{p_F} &=& \int_{0}^{L}dz \,{\cal K}_F(z) A_{p_F}(z) ,
 \nonumber \\
I_{p_B} &=& \int_{0}^{L}dz \,{\cal K}_B(z) A_{p_B}(z) .
\label{23}
\end{eqnarray}
The expressions contained in Eq. (\ref{22}) are correct
up to the second power of constants $ K_s $, $ K_i $, $ K_F $, and
$ K_B $. They show that principal squeeze
variances of single modes are larger than one and thus no
squeezing can occur in single modes.

The following expressions for the compound modes are reached
under the same assumptions:
\begin{eqnarray}   
 \lambda_{s_F,s_B} &=& 2 \left[ 1+ |I_{p_F}|^2 + |I_{p_B}|^2 \right] ,
  \nonumber \\
 \lambda_{s_F,i_F} &=& 2 \left[ 1- 2 |I_{p_F}| + 2 |I_{p_F}|^2 \right] ,
   \nonumber \\
 \lambda_{s_F,i_B} &=& 2 \left[ 1 + |I_{p_F}|^2 + |I_{p_B}|^2
  - 2 | I_{i,p_F} - I_{p_B,s} | \right] , \nonumber \\
 \lambda_{s_B,i_B} &=& 2 \left[ 1- 2 |I_{p_B}| + 2 |I_{p_B}|^2 \right] ,
   \nonumber \\
 \lambda_{s_F,p_F} &=& 2 \left[ 1 + |I_{p_F}|^2 - 2 | I_{s_F,p_F}|
   \right] , \nonumber \\
 \lambda_{s_F,p_B} &=& 2 \left[ 1 + |I_{p_F}|^2 \right] , \nonumber \\
 \lambda_{s_B,p_B} &=& 2 \left[ 1 + |I_{p_B}|^2 - 2 | I_{p_B,s_B} |
  \right] , \nonumber \\
 \lambda_{s_B,p_F} &=& 2 \left[ 1 + |I_{p_B}|^2 \right] , \nonumber \\
 \lambda_{p_F,p_B } &=& 2 ,
\label{24}
\end{eqnarray}
where
\begin{eqnarray}   
I_{i,p_F} &=& \int_{0}^{L} dz \int_{0}^{z}
   dz' {\cal K}^*_i(z) {\cal K}_F(z') A_{p_F}(z') ,
   \nonumber \\
I_{p_B,s} &=& \int_{0}^{L} dz \int_{0}^{z}
   dz' {\cal K}_B(z) A_{p_B}(z) {\cal K}_s(z') ,
   \nonumber \\
I_{s_F,p_F} &=& \int_{0}^{L} dz \int_{0}^{z}
   dz' {\cal K}^*_F(z) A_{s_F}(z) {\cal K}_F(z') A_{p_F}(z'),
   \nonumber \\
I_{p_B,s_B} &=& \int_{0}^{L} dz \int_{0}^{z}
   dz' {\cal K}_B(z) A_{p_B}(z) {\cal K}^*_B(z') A_{s_B}(z') .
 \nonumber \\
 & &
  \label{25}
\end{eqnarray}
According to Eq. (\ref{24}) squeezing can occur in compound modes
($ s_F, i_F $) and ($ s_B, i_B $) owing to nonlinear process
among the forward-propagating modes and that among the
backward-propagating modes. Light in compound mode
($ s_F, i_B $) can be squeezed if the linear coupling (describing
scattering of light in the photonic-band-gap waveguide) is
stronger than the
nonlinear one. As is seen from the
expression for $ \lambda_{s_F,i_B} $ in Eq. (\ref{24}) squeezing
originates in the nonlinear process and linear coupling between
forward- and backward-propagating fields is inevitable for
`transfer of squeezing' into this mode.
If classical amplitudes of the forward- (backward-) propagating
signal mode are greater than those of the forward- (backward-) propagating
pump mode, squeezing can be reached in compound mode ($ s_F,p_F $)
[($ s_B,p_B $)].

\subsection{Numerical analysis of squeezing}

Complete analysis of the behaviour of interacting modes
can be reached only numerically.

We assume a strong incident forward-propagating pump field and
also nonzero incident
signal and idler forward-propagating fields.
Squeezed light cannot be generated in single modes
in this case. However, compound modes ($ s_F,i_F $),
($ s_B,i_B $), and ($ s_F,i_B $) provide squeezed light at the
output under suitably chosen values of the waveguide parameters.

Values of parameters characterizing a real nonlinear planar
photonic-band-gap waveguide are assumed to lie around the point
given by
$ L = 2 $~mm, $ K_s = K_i = 5 $~mm$ {}^{-1} $,
$ K_F = K_B = 5 \times 10^{-7} $~mm$ {}^{-1} $ mV$ {}^{-1} $,
and $ A_{p_F} = 10^7 $ Vm$ {}^{-1} $.
That is why we concentrate our attention on quantitative analysis
of behaviour of the waveguide in the vicinity of this point.
We note that equality of values of linear coupling
constants $ K_s $ and $ K_i $ can barely be found in a real
waveguide but real modes can be chosen in such a way that values
of these constants are nearly the same.
The quantities
1 mm and $ 10^6 $~Vm$ {}^{-1} $ are used as appropriate units
in the following graphs.

\begin{figure}    
 {\raisebox{4 cm}{a)} \hspace{5mm}
 \resizebox{0.7\hsize}{!}{\includegraphics{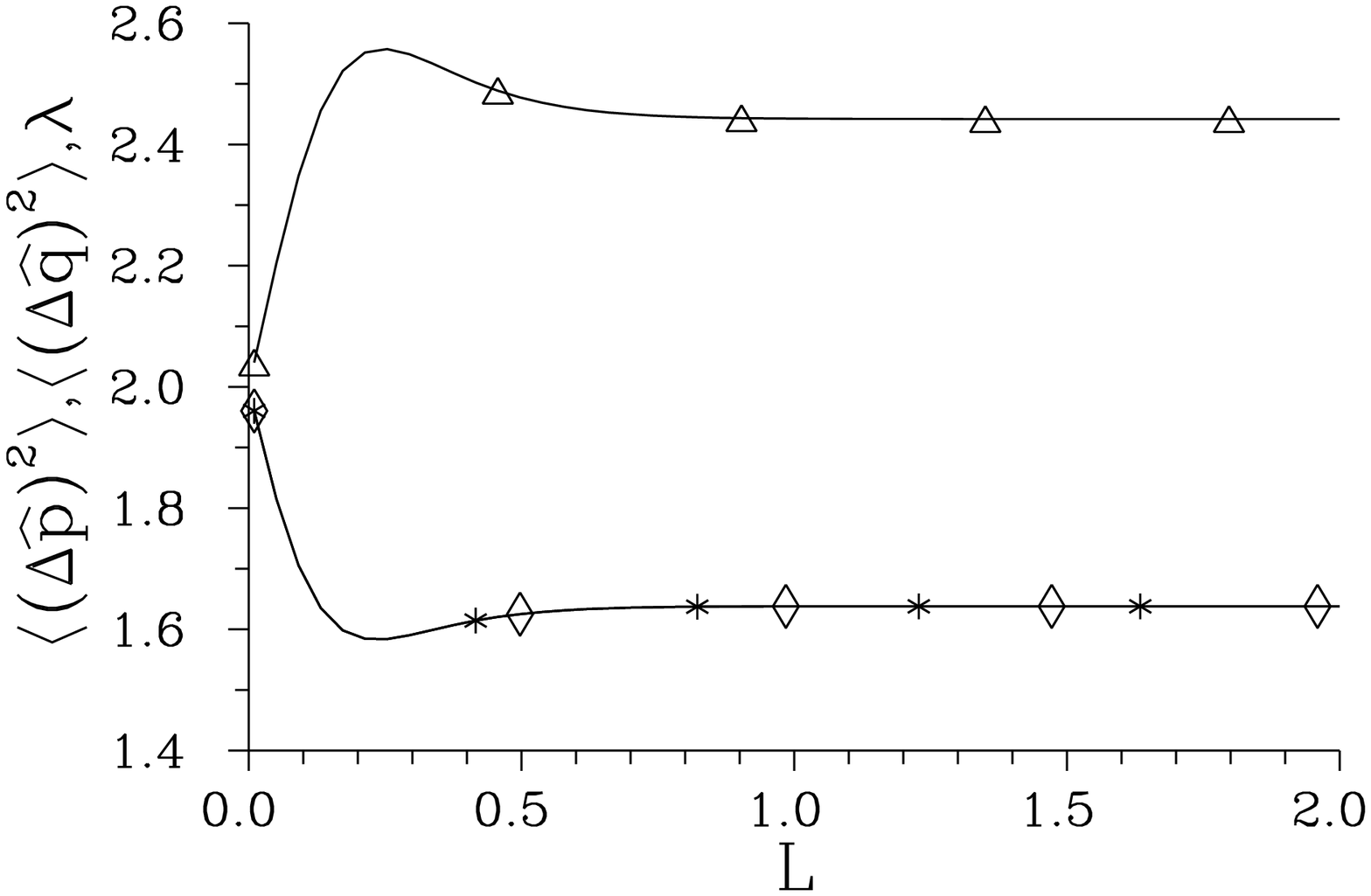}}

 \vspace{5mm}
 \raisebox{4 cm}{b)} \hspace{5mm}
 \resizebox{0.7\hsize}{!}{\includegraphics{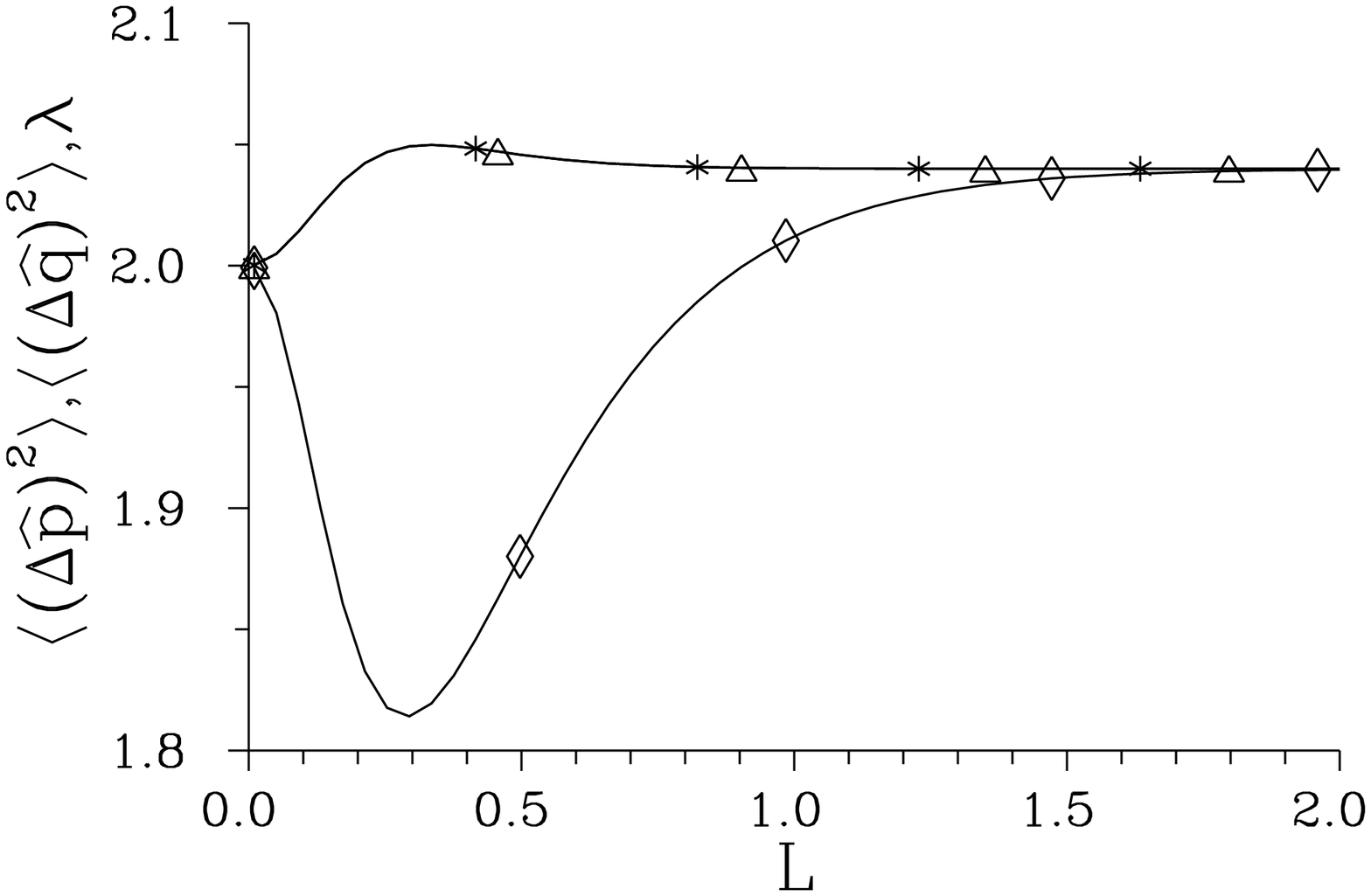}}

 \vspace{5mm}
 \raisebox{4 cm}{c)} \hspace{5mm}
 \resizebox{0.7\hsize}{!}{\includegraphics{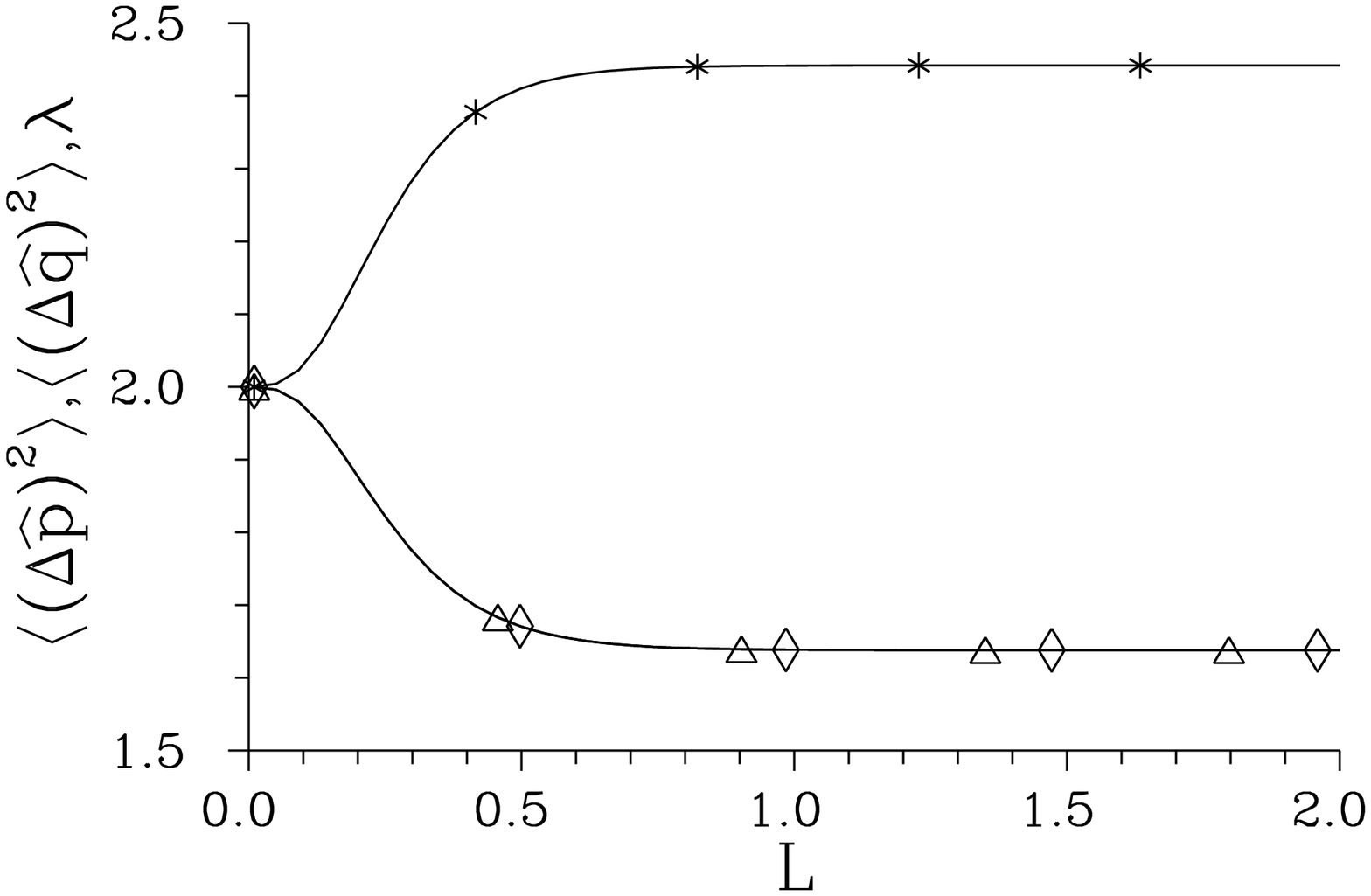}}}
 \vspace{2mm}

 \caption{Variances $ \langle (\Delta \hat{q})^2\rangle $ (solid
 line with $ \triangle $) and $ \langle (\Delta \hat{p})^2\rangle $
 ($ \star $) and principal squeeze variance $ \lambda $
 ($ \diamond $) for mode ($ s_F,i_F $) (a), ($ s_F,i_B $) (b),
 and ($ s_B,i_B $) (c) in dependence on length $ L $ of the
 structure; $ K_F = K_B = 5 \times 10^{-2} $,
 $ K_s = K_i = 5 $, $ \delta_s = \delta_i = 0 $,
 $ \delta_F = \delta_B = 0 $, $ A_{p_F} = 10 $,
 $ A_{s_F} = A_{i_F} = 0.1 $, $ A_{p_B} = A_{s_B} = A_{i_B} =0 $,
 incident coherent states are assumed. Backward-propagating fields exit
 the waveguide at $ z=0 $.}
 \label{fig1}
\end{figure}

The role of length $ L $ of the waveguide to squeezed-light
generation is revealed in Fig.~\ref{fig1}. We would like to remind that
any change of length $ L $ of the waveguide is accompanied by
a little change of
the period of corrugation, in order to maintain the same mode
profile and the same longitudinal confinement properties of the
modes. Fixing these criteria, the modes ($ s_F,i_F $) and ($
s_B,i_B $) are squeezed for any length $ L $ of the structure (see Figs.
\ref{fig1}a,c). However, the maximum value of squeezing is reached for
values of length $ L $ around 0.5 and then there is saturation in
values of squeezing. Squeezing of these modes around 20~\% can be
reached. The optimum value of length $ L $ lies around 0.3 for
mode ($ s_F,i_B $) composed of one forward- and one
backward-propagating field, as is shown in Fig.~\ref{fig1}b.
Squeezing in mode ($ s_F,i_B $) can occur owing to `scattering of
the already generated squeezed light' between forward- and
backward-propagating fields. Saturation of the behaviour for larger
values of length $ L $ is typical for interactions containing both
forward- and backward-propagating fields \cite{PerinaJr2000}.

\begin{figure}    
 \resizebox{0.8\hsize}{0.6\hsize}{\includegraphics{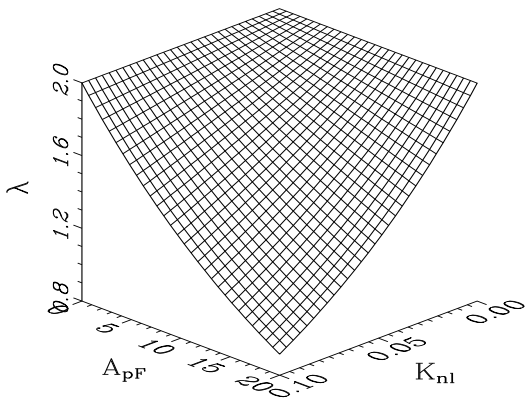}}

 \caption{Principal squeeze variance $ \lambda $
 of mode ($ s_F,i_F $) in dependence on nonlinear coupling
 constant $ K_{nl} $ ($ K_{nl} = K_F = K_B $ and input pump-pulse
 amplitude $ A_{p_F}(0) $; $ L = 2 $ and values of the other
 parameters are the same as in Fig. 1.}
 \label{fig2}
\end{figure}

Values of squeeze variances monotonously decrease with increasing values
of nonlinear coefficients $ K_F $ and $ K_B $ and with increasing
incident pump-pulse amplitudes $ A_{p_F} $. This clearly shows that
the origin of nonclassical properties of outgoing fields lies in
the nonlinear three-mode interaction.
As the graph in Fig. \ref{fig2} indicates,
an increase of values of nonlinear coefficient $ K_{nl} $
($ K_{nl} = K_F = K_B $) and pump-pulse amplitude $ A_{p_F} $
two times with
respect to the values characterizing the `fixed working point' of the
waveguide leads to values of squeezing
around 50~\% for mode ($ s_F,i_F $).

\begin{figure}    
 \resizebox{0.7\hsize}{0.6\hsize}{\includegraphics{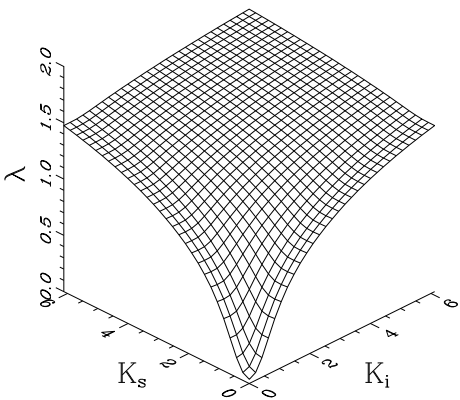}}

 \vspace{2mm}
 \caption{Principal squeeze variance $ \lambda $
 of mode ($ s_F,i_F $) in dependence on linear coupling
 constants $ K_s $ and $ K_i $; $ L = 2 $ and values of the other
 parameters are the same as in Fig. 1.}
 \label{fig3}
\end{figure}

Linear coupling between forward- and backward-propagating fields
(described by coupling constants $ K_s $ and $ K_i $)
originates in scattering of
forward-propagating fields in the photonic-band-gap waveguide
and does not lead to any nonclassical behaviour. Moreover,
a stronger linear coupling between signal modes and idler modes
suppresses squeezing, as is demonstrated in Fig. \ref{fig3}.
It holds for mode ($ s_F,i_F $) that the larger the values of linear
coupling constants the larger the values of quantities
characterizing
squeezing (see Fig. \ref{fig3}). However, nonzero values of
linear coupling
constants are necessary for obtaining squeezing in modes
($ s_F,i_B $) and ($ s_B,i_B $). The reason is that linear
coupling enables energy transfer to backward-propagating fields
(being in vacuum states at the input) from the
forward-propagating fields.

\begin{figure}    
 \resizebox{0.7\hsize}{!}{\includegraphics{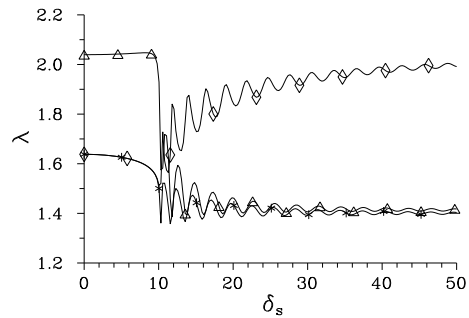}}

 \vspace{4mm}
 \caption{Principal squeeze variance $ \lambda $
 for modes ($ s_F,i_F $) (solid line with $ \star $),
 ($ s_F,i_B $) ($ \triangle $), and ($ s_B,i_B $) ($ \diamond $)
 in dependence on linear
 phase mismatch $ \delta_s $; $ L = 2 $ and values of the other
 parameters are the same as in Fig. 1.}
 \label{fig4}
\end{figure}

Phase matching of all optical-field interactions occurring in the
structure cannot be usually reached in a real structure.
A typical influence of nonzero linear phase mismatch is
demonstrated in Fig. \ref{fig4} describing behaviour of
modes ($ s_F,i_F $), ($ s_F,i_B $), and ($ s_B,i_B $).
A nonzero value of linear signal-fields phase mismatch $ \delta_s $
causes oscillations in quantities characterizing squeezing.
It effectively suppresses the influence of linear coupling
between the signal
modes and supports squeezed-light generation in mode
($ s_F,i_F $) this way.
On the other hand, values of squeeze variances for mode ($ s_B,i_B $)
increase with increasing values of phase mismatch
$ \delta_s $ because of lower values of amplitudes of the
backward-propagating pump mode $ p_B $.
Phase relations affected by nonzero values of $ \delta_s $
enable squeezed-light generation in mode ($ s_F,i_B $).

\begin{figure}    
 \raisebox{4 cm}{a)} \hspace{5mm}
 \resizebox{0.7\hsize}{!}{\includegraphics{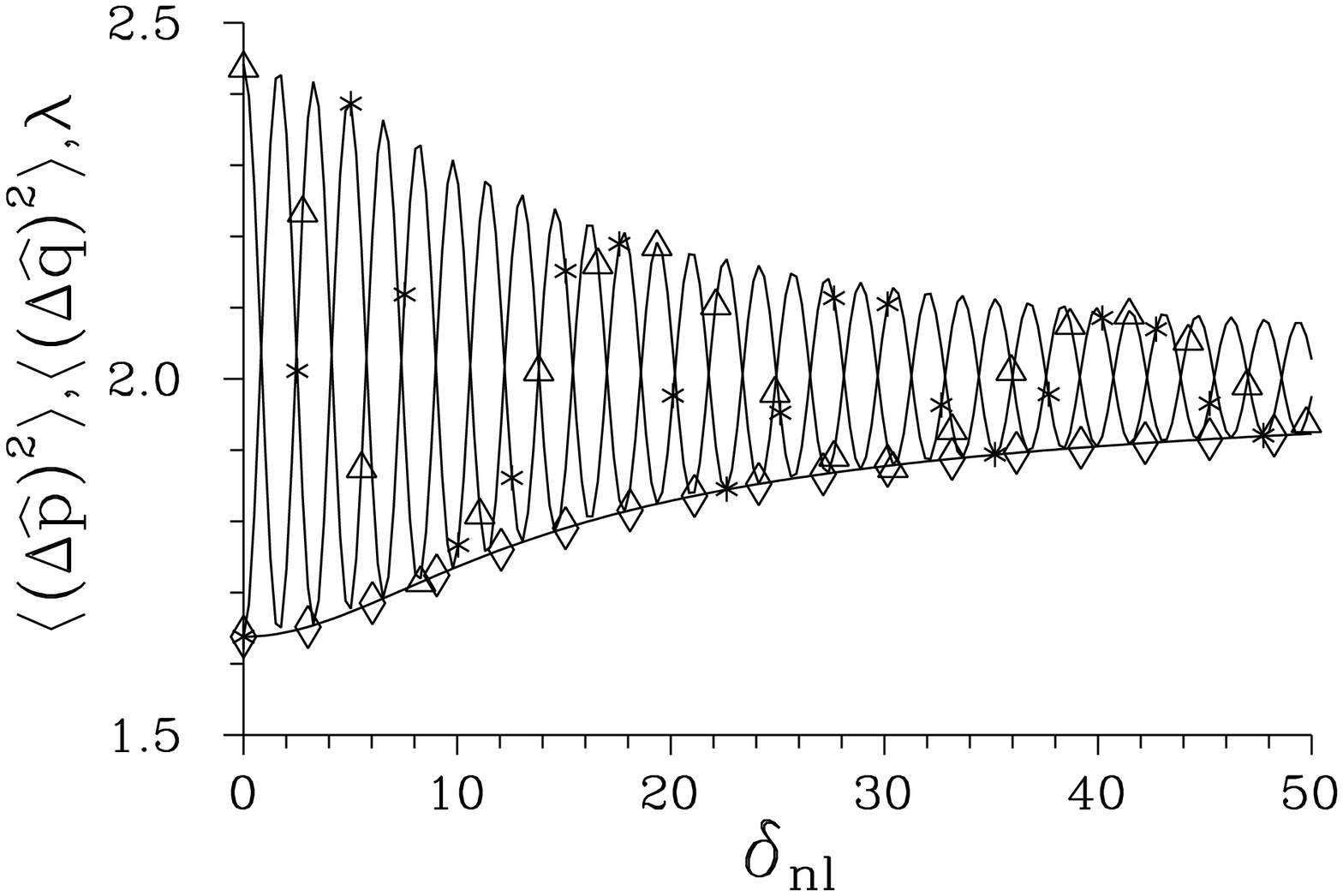}}

 \vspace{5mm}
 \raisebox{4 cm}{b)} \hspace{5mm}
 \resizebox{0.7\hsize}{!}{\includegraphics{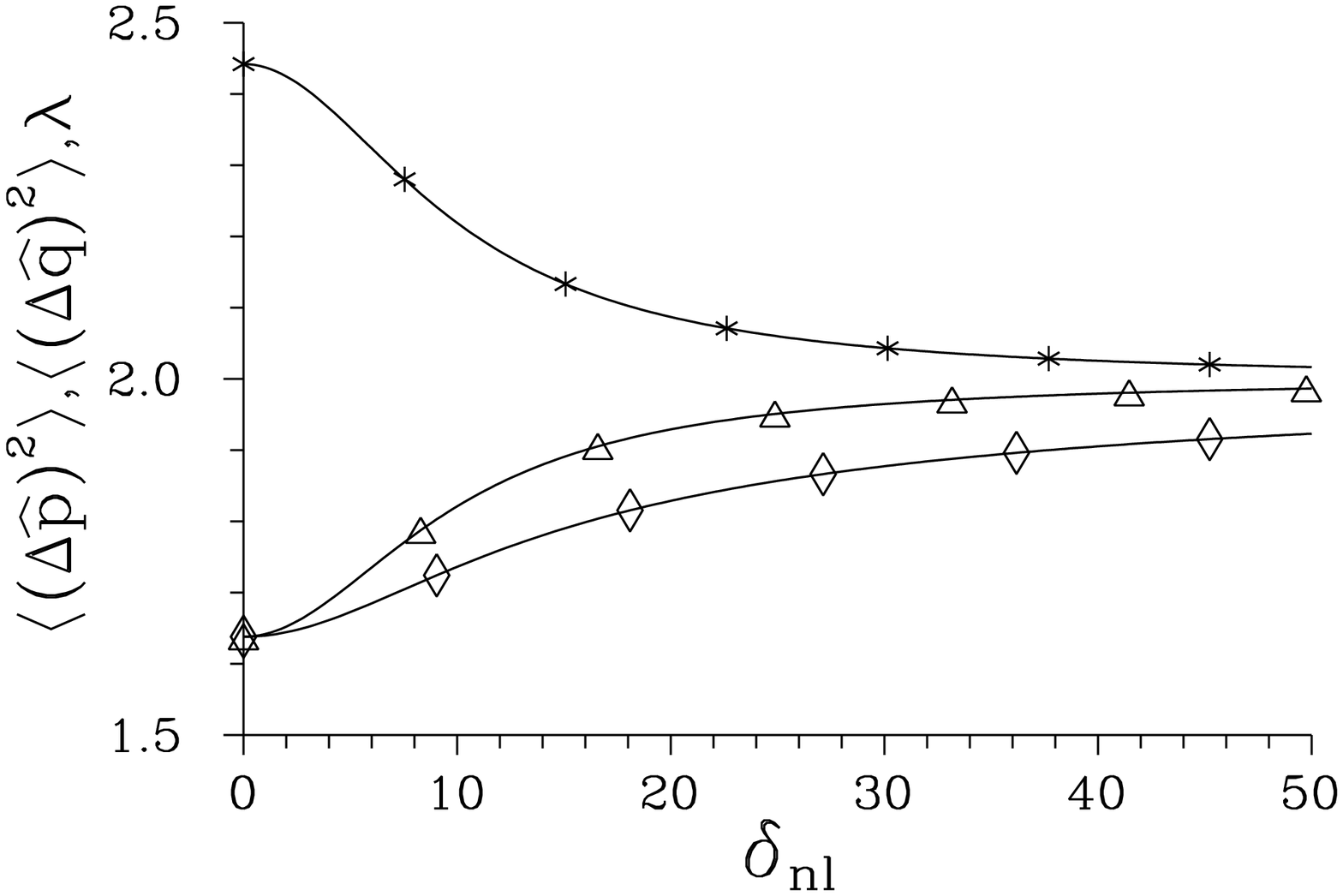}}
 \caption{Variances $ \langle (\Delta \hat{q})^2\rangle $ (solid
 line with $ \triangle $) and $ \langle (\Delta \hat{p})^2\rangle $
 ($ \star $) and principal squeeze variance $ \lambda $
 ($ \diamond $) for mode ($ s_F,i_F $) (a) and ($ s_B,i_B $) (b)
 in dependence on nonlinear phase mismatch $ \delta_{nl} $
 ($ \delta_{nl} = \delta_F = \delta_B $); $ L = 2 $ and
 values of the other parameters are the same as in Fig. 1.}
 \label{fig5}
\end{figure}

Increasing values of nonlinear phase mismatch
$ \delta_{nl} $ ($ \delta_{nl} = \delta_F = \delta_B $)
result in greater values of squeeze variances. This is demonstrated
in Fig. \ref{fig5} for modes ($ s_F,i_F $) and ($ s_B,i_B $).
This means that oscillations occurring along the waveguide
and having their origin in nonzero values of nonlinear phase
mismatch $ \delta_{nl} $ effectively lower
values of nonlinear coupling constants. Presence of such
oscillations is visible in the behaviour of variances of
quadratures $ \langle (\Delta \hat{q})^2\rangle $ and
$ \langle (\Delta \hat{p})^2\rangle $ in mode ($ s_F,i_F $)
(see Fig. \ref{fig5}a). The variances belonging to
mode ($ s_B,i_B $) do not show oscillations (see Fig.
\ref{fig5}b), because
oscillations are compensated in
nonlinear interaction among the backward-propagating fields.
Absence of such oscillations in
quantities characterizing squeezing in mode ($ s_B,i_B $)
indicates that squeezing in this mode has its origin only in
nonlinear interaction among the backward-propagating fields.

If we assume one of the forward-propagating signal and idler
fields in vacuum state at the input (parametric downconversion)
or both modes in vacuum states at the input (parametric
amplification) we observe qualitatively the same behaviour of
optical fields as discussed above.

\begin{figure}    
 \resizebox{0.7\hsize}{!}{\includegraphics{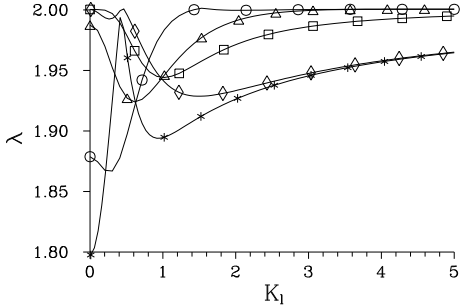}}

 \vspace{2mm}
 \caption{Principal squeeze variance $ \lambda $
 of mode ($ s_F,i_F $) (solid line with $ \star $),
 ($ s_F,i_B $) ($ \triangle $),
 ($ s_B,i_B $) ($ \diamond $), ($ s_F,p_F $) (circles), and
 ($ s_B,p_B $) (squares) in dependence on linear coupling
 constant $ K_l $ ($ K_l = K_s = K_i $); $ L = 2 $, $ A_{s_F} = 10 $,
 $ A_{i_F} = A_{p_F} = 1 $, values of the other parameters are the same as
 in Fig. 1.}
 \label{fig6}
\end{figure}

If the value of an incident amplitude of either the forward-propagating
signal field or forward-propagating idler field is greater than
that of the pump field (parametric upconversion)
squeezed light can be generated in compound modes ($ s_F,i_F $),
($ s_F,i_B $), and ($ s_B,i_B $). Moreover, squeezed light can
be reached also
in compound modes ($ s_F,p_F $) [($ i_F,p_F $)] and
($ s_B,p_B $) [($ i_B,p_B $)] assuming $ A_{s_F} > A_{p_F} $
[$ A_{i_F} > A_{p_F} $]. However, a strong negative influence of
linear coupling affects squeezed-light generation in this regime
(see Fig. \ref{fig6}).
We note that a measurement of squeezing in a compound mode
containing the pump field would require a more general scheme of
homodyne detection that would use two local oscillators, one
of them having the carrying frequency of the pump field, the second
one oscillating at signal- or idler-field frequency.

An incident squeezed light in the forward-propagating pump mode
does not support squeezed light generation. This means that values of
principal squeeze variances cannot be lower than
the minimum from the values characterizing the incident light
in a given mode and those obtained assuming an unsqueezed
incident forward-propagating pump mode.

\begin{figure}    
 \resizebox{0.6\hsize}{!}{\includegraphics{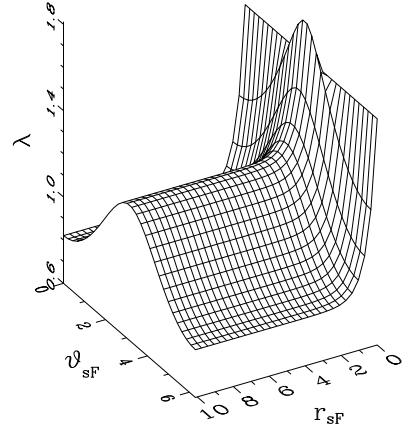}}

 \caption{Principal squeeze variance $ \lambda $
 of mode ($ s_B,i_B $) in dependence on squeeze parameter
 $ r_{sF} $ and squeeze phase $ \vartheta_{sF} $ characterizing
 the incident field;
 $ L = 2 $ and values of the other
 parameters are the same as in Fig. 1.}
 \label{fig7}
\end{figure}

If an incident squeezed light occurs in the forward-propagating signal
mode, smaller values of squeeze variances can be reached in modes
($ s_B,i_B $) and ($ s_B,i_F $) compared to those reached with
an unsqueezed input light. Values of the principle squeeze variances
then depend on incident squeeze parameter $ r $ and incident squeeze phase
$ \vartheta $ [see Fig. \ref{fig7} for mode ($ s_B,i_B $)].
Values of principle squeeze variances
of the other modes are greater than minimum from values
characterizing the incident light and those appropriate
for all unsqueezed incident beams.

\section{Sub-Poissonian-light generation}

Statistical properties of photoelectrons
emitted inside a detector reflect statistical properties
of integrated intensity $ \hat{W} $ of the radiation
impinging on the detector.
Integrated intensity $ \hat{W} $ of an optical field is defined as:
\begin{equation}    
 \hat{W} = \hat{A}^\dagger \hat{A} ,
\end{equation}
where $ \hat{A} $ means an electric-field-amplitude operator
of an optical field.
Detection process is characterized by normally-ordered moments of
integrated intensity $ \hat{W} $:
\begin{equation}     
 \langle W^k \rangle_{\cal N} = \langle \hat{A}^{\dagger k}
  \hat{A}^k \rangle , \hspace{10mm} k=2,3,\ldots
\end{equation}

Statistical properties of photoelectrons are usually judged
according to the value of Fano factor $ F_n $ defined as:
\begin{equation}     
 F_n = \frac{ \langle (\Delta n)^2 \rangle }{ \langle n \rangle}
 = 1 + \frac{ \langle (\Delta W)^2 \rangle_{\cal N} }{
  \langle W \rangle_{\cal N} } .
 \label{28}
\end{equation}
Symbol $ n $ denotes the number of photoelectrons,
$ \Delta n = n - \langle n \rangle $,  and
$ \Delta W = W - \langle W \rangle_{\cal N} $.
We assume that intensity operator $ \hat{W}_{ij} $ of the compound mode
($ i,j $) is determined using the relation
$ \hat{W}_{ij} = \hat{W}_i + \hat{W}_j $, where $ \hat{W}_i $
($ \hat{W}_j $) denotes intensity operator of mode $ i $
($ j $).
The condition $ F_n \ge 1 $ characterizes classical fields
whereas values of $ F_n $ lower than one can be reached only
for nonclassical fields (sub-Poissonian light). Such fields
have fluctuations in the number of photoelectrons suppressed
below the classical limit ($ F_n = 1 $ for a coherent state of laser
radiation).

In the following we pay attention to signal and idler fields that
are assumed to have intensities at single-photon level.
The planar nonlinear photonic-band-gap waveguide
is pumped by a strong forward-propagating pump field.
Classical amplitudes
$ A_{s_F} $, $ A_{i_F} $, $ A_{s_B} $, and $ A_{i_B} $
are zero in this case. Operator amplitudes $ \hat{A} $
of the signal and idler fields are then given just by
their linear operator corections $ \delta \hat{A} $.
The quantity 10 Vm$ {}^{-1} $ is used as a unit
for linear amplitude corrections. Mean values of intensities
using this unit are then directly equal to mean photon numbers.

\subsection{Weak-interaction approximation}

The expression for Fano factor $ F_n $ in Eq. (\ref{28}) shows
that sub-Poissonian light is generated
provided that $ \langle (\Delta W)^2 \rangle_{\cal N} < 0 $.
This condition cannot be fulfilled in a single-mode case as is indicated
by the following expressions valid in weak-interaction
approximation and assuming coherent states with
amplitudes $ \xi_j $ for incident quantum linear corrections
$ \delta \hat{A}_j $:
\begin{eqnarray}   
 \langle (\Delta W_{s_F})^2\rangle_{\cal N} &=& 2
  |I_{p_F}|^2 |\xi_{s_F}|^2 ,
  \nonumber \\
 \langle (\Delta W_{s_B})^2\rangle_{\cal N} &=& 2
  |I_{p_B}|^2 |\xi_{s_B}|^2 .
\end{eqnarray}
Constants $ I_{p_F} $ and $ I_{p_B} $ are defined in Eq.
(\ref{23}). However, we arrive at
the following expressions for compound modes under the same
conditions:
\begin{eqnarray}      
  \langle (\Delta (W_{s_F} +
    W_{i_F}))^2\rangle_{\cal N} &=&
    4{\rm Re} \left\{ I_{p_F} \xi^*_{s_F} \xi^*_{i_F} \right\}
    \nonumber \\
   & & \hspace{-3cm} \mbox{} + 2|I_{p_F}|^2
    \left( 1 + 3 |\xi_{s_F}|^2 + 3 |\xi_{i_F}|^2
      \right) \nonumber \\
   & & \hspace{-3cm} \mbox{} + 4 {\rm Re}
     \left\{ I_{p_F} I^*_s \xi^*_{s_B} \xi^*_{i_F} +
    I_{p_F} I_i \xi^*_{i_B} \xi^*_{s_F} \right\},
        \nonumber \\
 \langle (\Delta (W_{s_F}
   + W_{s_B}))^2\rangle_{\cal N} &=&  \nonumber \\
   & & \hspace{-3cm}
   2 (|I_{p_F}|^2 |\xi_{s_F}|^2 + |I_{p_B}|^2 |\xi_{s_B}|^2 ) ,
   \nonumber \\
  \langle (\Delta (W_{s_F}
   + W_{i_B}))^2\rangle_{\cal N} &=& \nonumber \\
   & & \hspace{-3cm}
     2(|I_{p_F}|^2 |\xi_{s_F}|^2 + |I_{p_B}|^2 |\xi_{i_B}|^2)
     \nonumber \\
   & & \hspace{-3cm} \mbox{} + 4
     {\rm Re} \left\{ -I_{i,p_F} \xi^*_{s_F} \xi^*_{i_B} +
      I_{p_B,s} \xi^*_{s_F} \xi^*_{i_B} \right\} ,
    \nonumber \\
   \langle (\Delta (W_{s_B} +
    W_{i_B}))^2\rangle_{\cal N} &=&
    4{\rm Re} \left\{ I_{p_B} \xi^*_{s_B} \xi^*_{i_B} \right\}
    \nonumber \\
    & & \hspace{-3cm} \mbox{} + 2|I_{p_B}|^2
      \left( 1 + 3 |\xi_{s_B}|^2 + 3 |\xi_{i_B}|^2 \right)
      \nonumber \\
    & &
    \hspace{-3cm} \mbox{}+ 4
      {\rm Re} \left\{ -I_{p_B} I_s \xi^*_{s_F} \xi^*_{i_B} -
      I_{p_B} I^*_i \xi^*_{i_F} \xi^*_{s_B} \right\} ;
     \nonumber \\
     & &
 \label{30}
\end{eqnarray}
\begin{eqnarray}     
 I_s &=& \int_{0}^{L} dz K_s(z) , \nonumber\\
 I_i &=& \int_{0}^{L} dz K^*_i(z) ;
\end{eqnarray}
$ I_{i,p_F} $ and $ I_{p_B,s} $ are defined in Eq. (\ref{25}).
Analysis of expressions in Eq. (\ref{30}) leads to the conclusion
that negative variances $ \langle (\Delta W)^2
\rangle_{\cal N} $ of
integrated intensity can occur in modes ($ s_F,i_F $) and
($ s_B,i_B $) provided that phases of the interacting fields are suitably
chosen. The best conditions for sub-Poissonian-light
generation in mode ($ s_F,i_F $) [($ s_B,i_B $)] occur provided
that $ \arg ( I_{p_F} \xi^*_{s_F} \xi^*_{i_F}) = \pi + 2\pi l $
[$ \arg (I_{p_B} \xi^*_{s_B} \xi^*_{i_B}) = \pi + 2\pi l $], $ l \in N $.
The occurrence of
negative variances $ \langle (\Delta W)^2 \rangle_{\cal N} $ of
integrated intensity in mode ($ s_F,i_B $) requires a stronger
linear interaction than the nonlinear one and again a suitably chosen
phases of the interacting optical fields as can be deduced
from the corresponding expression in Eq. (\ref{30}).

\subsection{Numerical analysis of sub-Poissonian-light
generation}

Complete analysis of sub-Poissonian behaviour of the interacting modes
can be done only numerically. Provided that states of the modes
can be
described in the framework of generalized superposition of signal and
noise \cite{Perina1991}, normally ordered moments of integrated
intensity as well as photon-number distribution can be determined
from parameters characterizing the state in terms of Laguerre
polynomials. Details can be found in
\cite{Perina1991,PerinaJr2000,Perinova1981}.

\begin{figure}    
{}\raisebox{4 cm}{a)} \hspace{5mm}
 \resizebox{0.5\hsize}{!}{\includegraphics{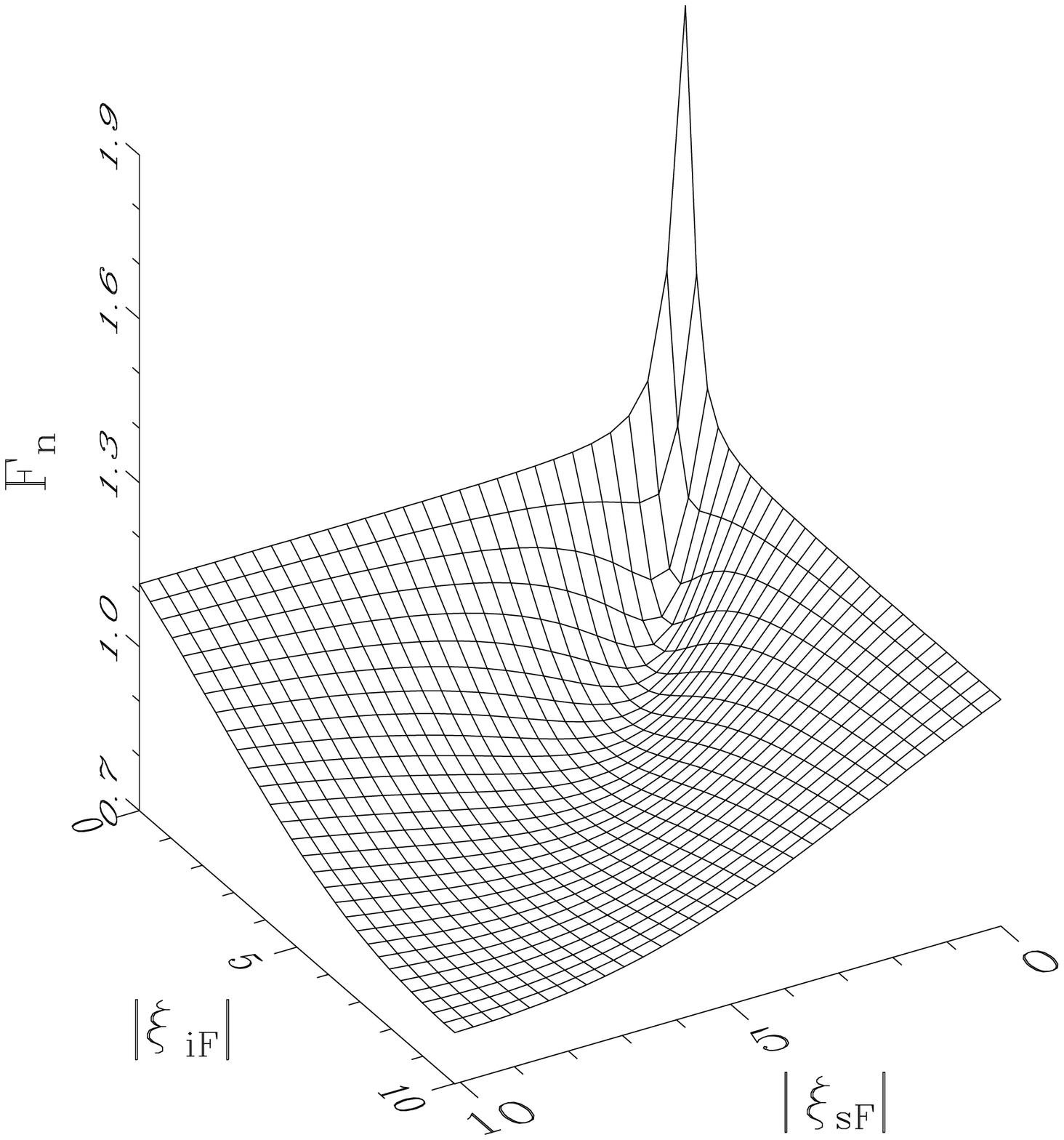}}

 \vspace{5mm}
 \raisebox{4 cm}{b)} \hspace{5mm}
 \resizebox{0.5\hsize}{!}{\includegraphics{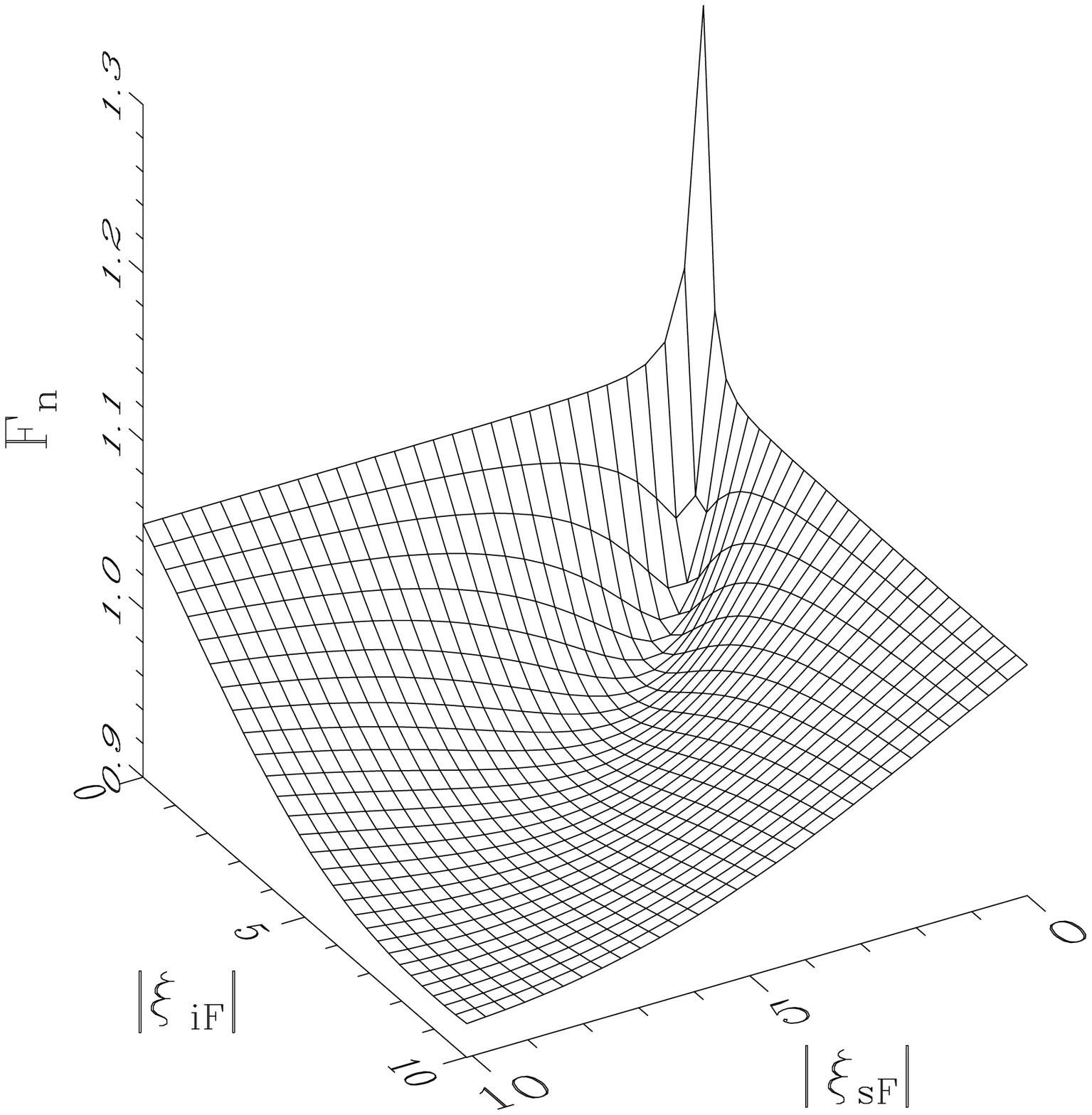}}

 \vspace{5mm}
 \raisebox{4 cm}{c)} \hspace{5mm}
 \resizebox{0.5\hsize}{!}{\includegraphics{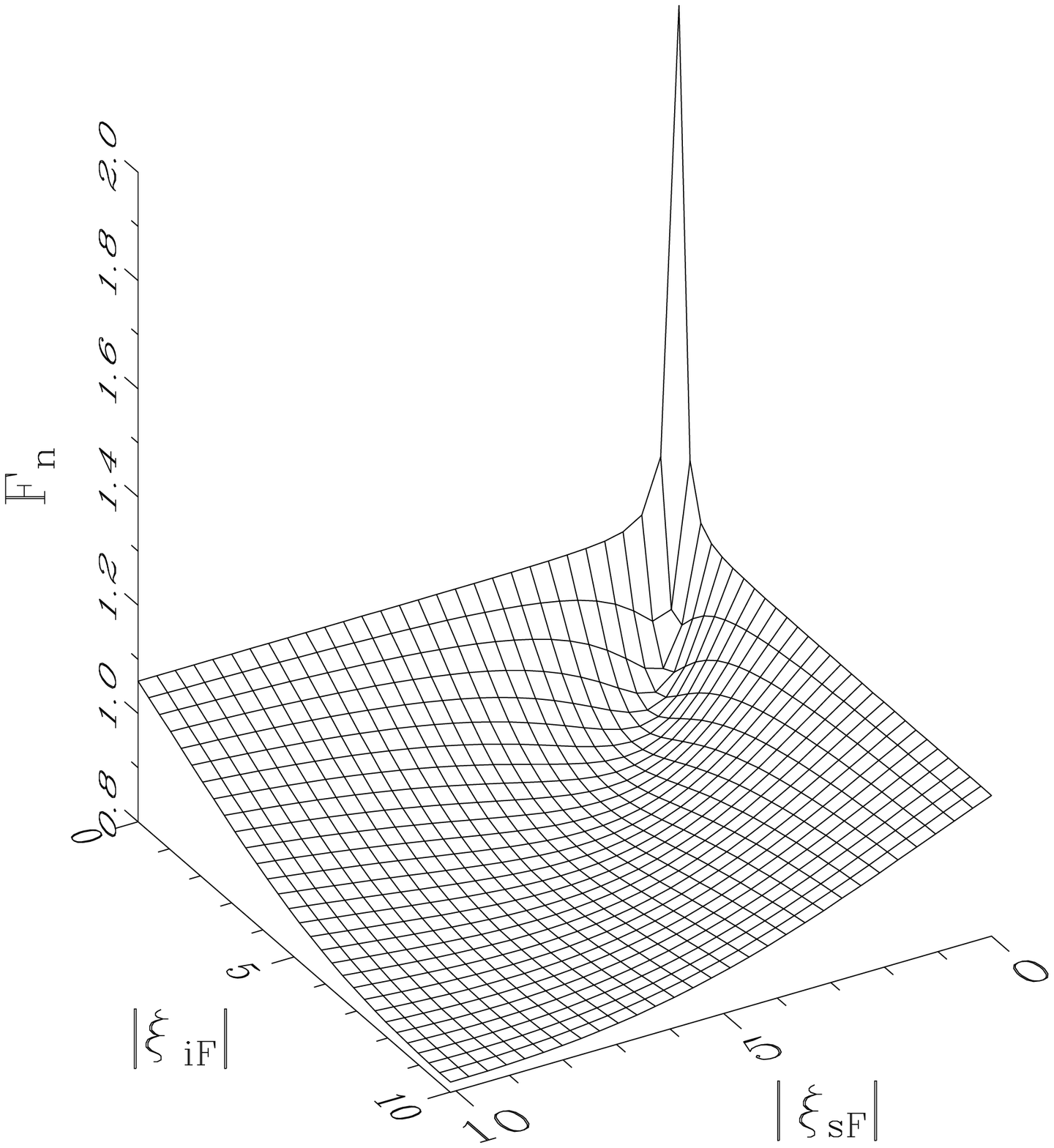}}

 \caption{Fano factor $ F_n $ for mode ($ s_F,i_F $) (a),
  ($ s_F,i_B $) (b),
  and ($ s_B,i_B $) (c) in dependence on amplitudes
  $ \xi_{s_F} $ and $ \xi_{i_F} $;
  linear operator amplitude corrections $ \delta \hat{A} $
  are assumed to be in coherent states
  with amplitudes $ \xi $ at the input;
  $ K_F = K_B = 5 \times 10^{-2} $,
  $ K_s = K_i = 5 $, $ L=0.2 $, $ \delta_s = \delta_i = 0 $,
  $ \delta_F = \delta_B = 0 $, $ A_{p_F} = 10 $,
  $ A_{s_F} = A_{i_F} = 0 $, $ A_{p_B} = A_{s_B} = A_{i_B} =0 $;
  $ \arg (\xi_{s_F}) = \arg (\xi_{i_F}) = 0 $,
  $ \xi_{p_F} = \xi_{s_B} = \xi_{i_B} = \xi_{p_B} = 0 $.}
 \label{fig8}
\end{figure}

Numerical analysis has shown that sub-Poissonian light can be generated
only in compound modes ($ s_F,i_F $), ($ s_F,i_B $), and
($ s_B,i_B $). Moreover, sub-Poissonian-light generation requires
nonzero values of the incident weak forward-propagating signal and
idler fields. As is demonstrated in Fig. \ref{fig8}, even fields
containing less than one photon on average are sufficient to stimulate
generation of sub-Poissonian light. Then we can have light with
values of Fano factor $ F_n $ around 0.8 in mode ($ s_F,i_F $).
Scattering of signal and idler fields leads to sub-Poissonian
statistics in modes ($ s_F,i_B $) and ($ s_B,i_B $).
This is remarkable especially for mode ($ s_B,i_B $) being
in vacuum state at the
input (see Fig. \ref{fig8}c). The larger the incident signal- and
idler-field amplitudes, the smaller the values of Fano factor
$ F_n $.

\begin{figure}    
 \resizebox{0.7\hsize}{!}{\includegraphics{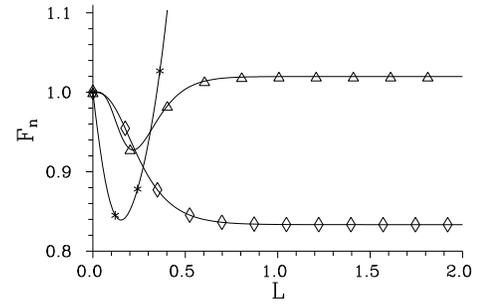}}

 \caption{Fano factor $ F_n $ of mode ($ s_F,i_F $)
  (solid line with $ \star $), ($ s_F,i_B $)
  ($ \triangle $), and ($ s_B,i_B $) ($ \diamond $) as a function
  of length $ L $;
  $ \xi_{s_F} = -1 $, $ \xi_{i_F} = 1 $, and values of the other
  parameters are the same as in Fig. 8.}
 \label{fig9}
\end{figure}

The possibility to generate sub-Poissonian light depends on length
$ L $ of the waveguide (see Fig. \ref{fig9}). Sub-Poissonian
statistics of light in modes ($ s_F,i_F $) and ($ s_F,i_B $) occur
only for shorter waveguides. On the other hand, greater values of
length $ L $ are convenient for sub-Poissonian statistics of light
in mode ($ s_B,i_B $). Quantities characterizing this mode
composed of only backward-propagating fields show saturation in
dependence on length $ L $ being typical for systems containing
backward-propagating fields.

\begin{figure}    
 \resizebox{0.7\hsize}{!}{\includegraphics{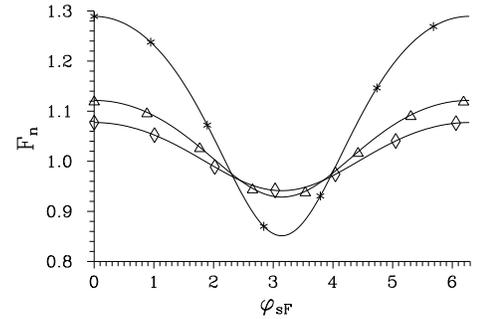}}

 \caption{Fano factor $ F_n $ of mode ($ s_F,i_F $)
  (solid line with $ \star $), ($ s_F,i_B $)
  ($ \triangle $), and ($ s_B,i_B $) ($ \diamond $) in dependence
  on the phase $ \varphi_{s_F} $ [$ \varphi_{s_F} = \arg (\xi_{s_F}) $]
  in units of $ \pi $;
  $ |\xi_{s_F}| = 1 $, $ \xi_{i_F} = 1 $, and values of the other
  parameters are the same as in Fig. 8.}
 \label{fig10}
\end{figure}

As is indicated by analytical expressions in Eq. (\ref{30}) valid in
weak-interaction approximation an efficient generation of
sub-Poissonian light requires a suitable choice of phases of
the forward-propagating fields.
A strong influence of phase $ \varphi_{s_F} $
of the incident forward-propagating signal field on values of
Fano factor $ F_n $ of modes ($ s_F,i_F $), ($ s_F,i_B $),
and ($ s_B,i_B $) is visualized in Fig. \ref{fig10}.

\begin{figure}    
 \resizebox{0.7\hsize}{!}{\includegraphics{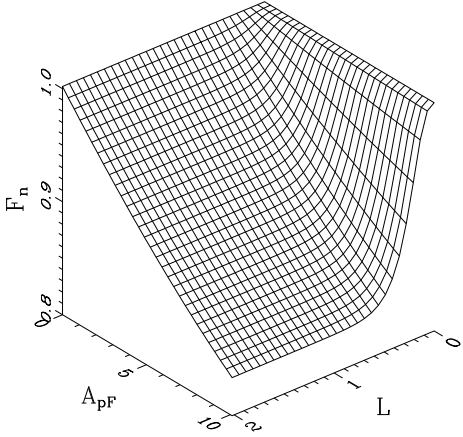}}

 \caption{Fano factor $ F_n $ of mode ($ s_B,i_B $) in dependence
  on incident pump amplitude $ A_{p_F} $ and length $ L $;
  $ \xi_{s_F} = -1 $, $ \xi_{i_F} = 1 $, and values of the other
  parameters are the same as in Fig. 8.}
 \label{fig11}
\end{figure}

In order to reach smaller values of Fano factor $ F_n $ greater
values of the pump field $ A_{p_F} $ are necessary. The larger the
pump amplitude $ A_{p_F} $, the smaller the values of
Fano factor $ F_n $. For example, assuming mode ($ s_B,i_B $)
$ F_n \approx 0.9 $ for $ A_{p_F} = 5 $, whereas
$ F_n \approx 0.8 $ for $ A_{p_F} = 10 $
(see Fig. \ref{fig11}).

\begin{figure}    
 \resizebox{0.7\hsize}{!}{\includegraphics{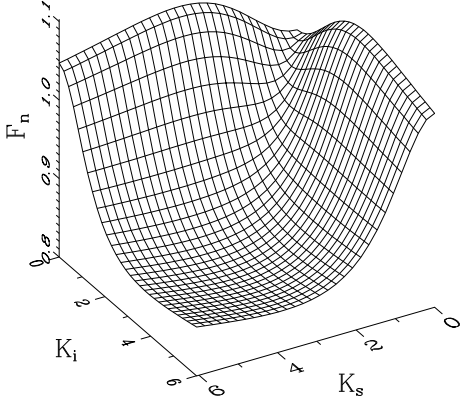}}

 \caption{Fano factor $ F_n $ of mode ($ s_B,i_B $) in dependence
  on linear coupling constants $ K_s $ and $ K_i $;
  $ L = 0.5 $, $ \xi_{s_F} = -1 $, $ \xi_{i_F} = 1 $,
  and values of the other
  parameters are the same as in Fig. 8.}
 \label{fig12}
\end{figure}

Linear coupling of modes
influences sub-Poissonian-light generation as follows.
The greater the values of linear coupling constants
$ K_s $ and $ K_i $, the greater values of Fano factor
$ F_n $ are reached in mode ($ s_F,i_F $). However, even larger
values of $ K_s $ and $ K_i $ enable to generate light with
sub-Poissonian statistics.
In mode ($ s_F,i_B $), greater values of $ K_s $ result in
greater values of $ F_n $. On the other hand, greater values of
$ K_i $ lead to lower values of $ F_n $ owing to a stronger
scattering between the forward- and backward-propagating
idler modes.
The behaviour of mode ($ s_B,i_B $) is shown in Fig. \ref{fig12}.
If strength of scattering between the signal and idler fields
is more-less balanced ($ K_s \approx K_i $), larger values of
linear coupling constants $ K_s $ and $ K_i $
provide smaller values of Fano factor
$ F_n $. Otherwise greater values of $ K_s $ and $ K_i $ neednot
necessarily lead to smaller values of Fano factor $ F_n $.

\begin{figure}    
 \resizebox{0.7\hsize}{!}{\includegraphics{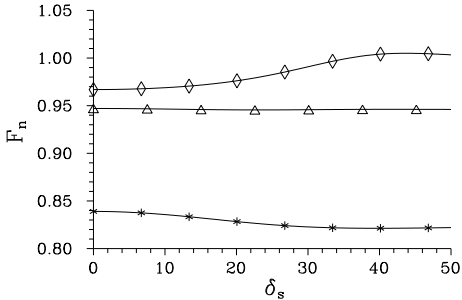}}

 \caption{Fano factor $ F_n $ of mode ($ s_F,i_F $)
  (solid line with $ \star $), ($ s_F,i_B $)
  ($ \triangle $), and ($ s_B,i_B $) ($ \diamond $) in dependence
  on linear phase mismatch $ \delta_s $;
  $ \xi_{s_F} = -1 $, $ \xi_{i_F} = 1 $, and values of the other
  parameters are the same as in Fig. 8.}
 \label{fig13}
\end{figure}

Complete phase matching of all interactions cannot be usually
reached in real nonlinear photonic-band-gap waveguides.
The effect of linear signal-field phase mismatch $ \delta_s $ on
values of Fano factor $ F_n $ is relatively weak
(see Fig. \ref{fig13}).
Greater values of
$ \delta_s $ decrease values of Fano factor $ F_n $ of mode
($ s_F,i_F $) but they increase values of Fano factor $ F_n $ of mode
($ s_B,i_B $). This behaviour can be understood claiming that
nonzero values of linear phase mismatch $ \delta_s $ effectively
weaken the linear coupling constant between the signal modes.

\begin{figure}    
 \resizebox{0.7\hsize}{!}{\includegraphics{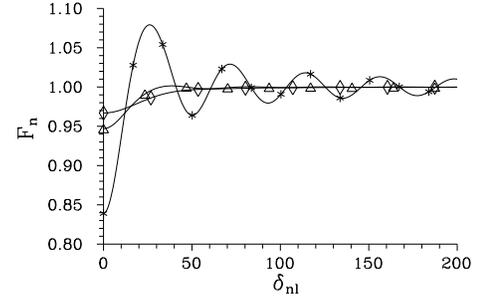}}

 \caption{Fano factor $ F_n $ of mode ($ s_F,i_F $)
  (solid line with $ \star $), ($ s_F,i_B $)
  ($ \triangle $), and ($ s_B,i_B $) ($ \diamond $) in dependence
  on nonlinear phase mismatch $ \delta_{nl} $
  ($ \delta_{nl} = \delta_F = \delta_B $);
  $ \xi_{s_F} = -1 $, $ \xi_{i_F} = 1 $, and values of the other
  parameters are the same as in Fig. 8.}
 \label{fig14}
\end{figure}

Nonzero values of nonlinear phase mismatches $ \delta_F $ and
$ \delta_B $ effectively weaken nonlinear interaction in a
waveguide and this acts against sub-Poissonian-light generation.
As is shown in Fig. \ref{fig14} sub-Poissonian light cannot be
generated for greater values of
$ \delta_{nl} $ ($ \delta_{nl} = \delta_F = \delta_B $)
at all.

\begin{figure}    
 \resizebox{0.7\hsize}{!}{\includegraphics{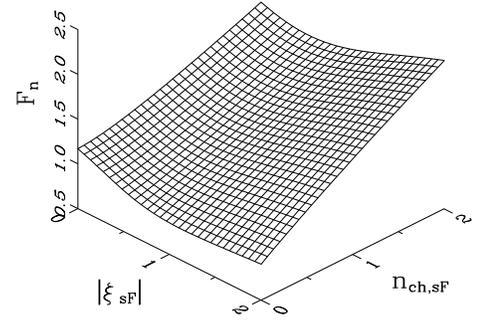}}

 \caption{Fano factor $ F_n $ of mode ($ s_F,i_F $) in dependence
  on coherent amplitude $ \xi_{s_F} $ and mean number of incident
  noisy photons $ n_{ch,s_F} $;
  $ \arg(\xi_{s_F}) = 0, $, $ \xi_{i_F} = -1 $,
  and values of the other
  parameters are the same as in Fig. 8.}
 \label{fig15}
\end{figure}

Noise present in incident fields leads to suppression of
sub-Poissonian behaviour. Generation of sub-Poissonian
light is rather sensitive to values of incident noise,
as is shown for mode ($ s_F,i_F $) in Fig. \ref{fig15}.

\begin{figure}    
 \resizebox{0.7\hsize}{!}{\includegraphics{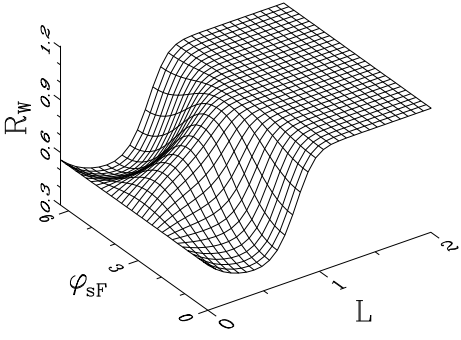}}

 \caption{Second reduced moment $ R_W $ of integrated intensity
  $ W $ of mode $ s_F $ in dependence on length $ L $ of the
  waveguide and incident phase $ \varphi_{s_F} $ [$ \varphi_{s_F}
  = \arg(\xi_{s_F}) $];
  $ |\xi_{s_F}| = 2, $, $ \xi_{i_F} = 2 $, $ n_{ch,s_F}=2 $,
  and values of the other
  parameters are the same as in Fig. 8.}
 \label{fig16}
\end{figure}

A nonlinear planar waveguide can be also used for the suppression of
incident noise. This effect has its origin in sensitivity of the
nonlinear process to the phase of an incident light. To be more
specific, the amplification coefficient of the incident light depends
on its initial phase. If the central phase of the incident field
(corresponding to a coherent signal amplitude) has the strongest
amplification then the noisy part (with a blurred phase) is
less amplified on average and signal-to-noise ratio increases.
Reduction of the incident noise is demonstrated in Fig. \ref{fig16}
for single mode $ s_F $ using second reduced moment $ R_W $
of integrated intensity $ W $ [$ R_W = \langle W^2
\rangle_{\cal N} / \langle W \rangle_{\cal N}^2 $].
Reduction of the incident noise occurs only for shorter lengths $ L $
and values of incident phases in a certain region (amplification of the
incident field has to occur in the waveguide).

\section{Conclusions}

We have analyzed optical parametric process occurring in a
nonlinear planar photonic-band-gap waveguide. It has been shown
that squeezed light and light with sub-Poissonian photon-number
statistics can be generated only in compound modes containing a)
signal and idler forward-propagating fields, b) signal (idler)
forward-propagating field and idler (signal) backward-propagating
field, and c) signal and idler backward-propagating fields. The role
of parameters characterizing a real nonlinear photonic-band-gap
waveguide in generation of light with nonclassical properties has
been analyzed in detail and understood. Also the influence of
incident squeezed light and incident noise has been elucidated.
The waveguide can be used for the suppression of incident noise.
The obtained analysis will be used for the construction of a real
nonlinear photonic-band-gap waveguide.

\acknowledgments{This work was supported by the COST project OC
P11.003 of the Czech Ministry of Education (M\v{S}MT) being part
of the ESF project COST P11 and by grant LN00A015 of the Czech
Ministry of Education. Support comming from cooperation
agreement between Palack\'{y} University and University
La Sapienza in Rome is acknowledged.}

\appendix

\section{Quantum derivation of the nonlinear equations}

Quantum description of nonlinearly interacting optical modes is
based on the construction of momentum operator $ \hat{G}(z) $
which determines Heisenberg equations of motion:
\begin{equation}   
 \frac{d\hat{X}}{dz} = - \frac{i}{\hbar} \left[ \hat{G}, \hat{X}
  \right] ;
\label{A1}
\end{equation}
$ \hat{X} $ stands for an arbitrary operator and $ [\;,\;] $
means a commutator.

If a nonlinear interaction involves counterpropagating beams, no
momentum operator can be straightforwardly assigned to
Heisenberg equations
as they are written in Eq.(\ref{6}). However, we can proceed as
follows \cite{PerinaJr2000}. We assume
a nonlinear interaction among all involved fields
as if they propagate in one direction and then we can write the momentum
operator $ \hat{G}(z) $ in the form:
\begin{eqnarray}  
 \hat{G}(z) = \sum_{a=s_F,i_F,p_F} \hbar (k_a)_z \hat{a}^\dagger_a
 \hat{a}_a  + \sum_{a=s_B,i_B,p_B} \hbar (k_a)_z \hat{a}^\dagger_a
 \hat{a}_a  \nonumber \\
 \mbox{} + \left[ \hbar K_{s} \exp(i\delta_l z)
  \hat{a}^\dagger_{s_F} \hat{a}_{s_B} + \hbar K_{i} \exp(i\delta_l z)
  \hat{a}^\dagger_{i_F} \hat{a}_{i_B} + {\rm h.c.} \right]
  \nonumber \\
 \mbox{} - \left[ 2i\hbar K_{F} \hat{a}_{p_F}\hat{a}^\dagger_{s_F}
  \hat{a}^\dagger_{i_F} + 2i\hbar K_{B}
  \hat{a}_{p_B}\hat{a}^\dagger_{s_B} \hat{a}^\dagger_{i_B} +
  {\rm h.c.} \right] . \nonumber \\
\end{eqnarray}
We now substitute the creation operators ($ \hat{a}^\dagger $) of
the backward-propagating fields by newly introduced
fictions annihilation
operators ($ \hat{b} $) and vice versa, i.e.
\begin{eqnarray}    
 \hat{a}^\dagger_{s_B} \leftarrow \hat{b}_{s_B} ,
 \hat{a}^\dagger_{i_B} \leftarrow \hat{b}_{i_B} ,
 \hat{a}^\dagger_{p_B} \leftarrow \hat{b}_{p_B} , \nonumber \\
 \hat{a}_{s_B} \leftarrow \hat{b}^\dagger_{s_B} ,
 \hat{a}_{i_B} \leftarrow \hat{b}^\dagger_{i_B} ,
 \hat{a}_{p_B} \leftarrow \hat{b}^\dagger_{p_B} .
\label{A3}
\end{eqnarray}
Heisenberg equations in Eq. (\ref{A1}) then have the form:
\begin{eqnarray}    
 \frac{d\hat{a}_{s_F}}{dz} &=& i (k_{s_F})_z \hat{a}_{s_F}
  + i K_{s}\exp(i\delta_l z) \hat{b}^\dagger_{s_B}
  \nonumber \\
 & & \mbox{} + 2K_{F}\hat{a}_{p_F}\hat{a}^\dagger_{i_F},
  \nonumber \\
 \frac{d\hat{a}_{i_F}}{dz} &=& i (k_{i_F})_z \hat{a}_{i_F}
  + i K_{i}\exp(i\delta_l z) \hat{b}^\dagger_{i_B}
  \nonumber \\
 & & \mbox{} + 2K_{F}\hat{a}_{p_F}\hat{a}^\dagger_{s_F},
  \nonumber \\
 \frac{d\hat{b}^\dagger_{s_B}}{dz} &=& -i (k_{s_B})_z
  \hat{b}^\dagger_{s_B} - i K^*_{s}\exp(-i\delta_l z)
   \hat{a}_{s_F}
  \nonumber \\
 & & \mbox{} - 2K_{B}\hat{b}^\dagger_{p_B}\hat{b}_{i_B},
  \nonumber \\
 \frac{d\hat{b}^\dagger_{i_B}}{dz} &=& -i (k_{i_B})_z
  \hat{b}^\dagger_{i_B} - i K^*_{i}\exp(-i\delta_l z)
   \hat{a}_{i_F}
  \nonumber \\
 & & \mbox{} - 2K_{B}\hat{b}^\dagger_{p_B}\hat{b}_{s_B},
  \nonumber \\
 \frac{d\hat{a}_{p_F}}{dz} &=& i (k_{p_F})_z
  \hat{a}_{p_F}
  - 2K^*_{F}\hat{a}_{s_F}\hat{a}_{i_F} ,
  \nonumber \\
 \frac{d\hat{b}^\dagger_{p_B}}{dz} &=& -i (k_{p_B})_z
  \hat{b}^\dagger_{p_B}
  + 2K^*_{B}\hat{b}^\dagger_{s_B}\hat{b}^\dagger_{i_B}.
\label{A4}
\end{eqnarray}
Returning to the original operators $ \hat{a}^\dagger $, $ \hat{a} $
in Eq. (\ref{A4}) using substitution in Eq. (\ref{A3}) and
transforming Eq. (\ref{A4}) into interaction picture
($ \hat{a}_a(z) = \hat{A}_a(z) \exp[i(k_a)_z z] $) we arrive at
the system of equations written in Eq. (\ref{6}). We only have
electric-field-amplitude operators instead of classical
electric-field amplitudes occurring in Eq. (\ref{6}).

Quantum interpretation of the nonlinear process of three mode
interaction at single-photon level then immediately provides
the following conservation law of the overall number of photons
in the interaction:
\begin{eqnarray}     
 \frac{d}{dz} \left\langle \hat{A}^\dagger_{s_F} \hat{A}_{s_F}
 + \hat{A}^\dagger_{i_F} \hat{A}_{i_F} + 2 \hat{A}^\dagger_{p_F}
  \hat{A}_{p_F} \right. \nonumber \\
  \mbox{} \left. - \hat{A}^\dagger_{s_B}\hat{A}_{s_B} -
  \hat{A}^\dagger_{i_B}\hat{A}_{i_B} -
  2 \hat{A}^\dagger_{p_B}\hat{A}_{p_B} \right\rangle = 0.
\end{eqnarray}
The symbol $ \langle \;\; \rangle $ denotes mean quantum statistical
value.

\section{Identities obeyed by a solution of Eq. (\ref{11})}

Solution of Eq. (\ref{11}) for the annihilation operators of
linear operator corrections can be expressed as:
\begin{eqnarray}   
 \pmatrix{ \delta \hat{A}(L) \cr \delta \hat{B}^\dagger(L) } &=&
 \pmatrix{ u_{11} & u_{12} \cr u_{21} & u_{22} }
  \pmatrix{ \delta \hat{A}(0) \cr \delta\hat{B}^\dagger(0) }
  \nonumber \\
 & & \mbox{} + \pmatrix{ v_{11} & v_{12} \cr v_{21} & v_{22} }
  \pmatrix{ \delta\hat{A}^\dagger(0) \cr \delta\hat{B}(0) }
\label{B1}
\end{eqnarray}
and
\begin{equation}    
 \delta \hat{A}(z) = \pmatrix{ \delta\hat{A}_{s_F}(z) \cr
  \delta\hat{A}_{i_F}(z) \cr \delta\hat{A}_{p_F}(z) } , \;\;
 \delta \hat{B}^\dagger(z) = \pmatrix{ \delta\hat{A}_{s_B}(z) \cr
  \delta\hat{A}_{i_B}(z) \cr \delta\hat{A}_{p_B}(z) } .
\end{equation}

Quantum method of the derivation of Eq. (\ref{6}) described in
Appendix A indicates that the following commutation relations
have to be fulfilled:
\begin{eqnarray}    
 [\delta\hat{A}_i(L),\delta\hat{A}_k(L)] &=& 0,
  \nonumber \\
 {} [\delta\hat{A}_i(L),\delta\hat{A}^\dagger_k(L)] &=& \delta_{ik},
  \nonumber \\
 {} [\delta\hat{A}_i(L),\delta\hat{B}_k(L)] &=& 0,
  \nonumber \\
 {} [\delta\hat{A}_i(L),\delta\hat{B}^\dagger_k(L)] &=& 0,
  \nonumber \\
 {} [\delta\hat{B}^\dagger_i(L),\delta\hat{B}_k(L)] &=& -\delta_{ik},
  \nonumber \\
 {} [\delta\hat{B}^\dagger_i(L),\delta\hat{B}^\dagger_k(L)] &=& 0.
\label{B3}
\end{eqnarray}

Substitution of expressions in Eq. (\ref{B1}) into relations
in Eq. (\ref{B3}) provides identities for the matrices $ u $
and $ v $ defined in Eq. (\ref{B1}):
\begin{eqnarray}     
 \sum_{j} \left[(u_{11})_{ij} (v_{11})_{kj} - (u_{12})_{ij} (v_{12})_{kj}
 - (v_{11})_{ij} (u_{11})_{kj} \right. \nonumber \\
 \mbox{} \left. + (v_{12})_{ij} (u_{12})_{kj}
 \right] = 0, \nonumber \\
 \sum_{j} \left[(u_{11})_{ij} (u^*_{11})_{kj} - (u_{12})_{ij}
 (u^*_{12})_{kj} - (v_{11})_{ij} (v^*_{11})_{kj} \right. \nonumber \\
 \mbox{} \left. + (v_{12})_{ij} (v^*_{12})_{kj}
 \right] = \delta_{ik}, \nonumber \\
 \sum_{j} \left[(u_{11})_{ij} (u^*_{21})_{kj} - (u_{12})_{ij}
 (u^*_{22})_{kj} - (v_{11})_{ij} (v^*_{21})_{kj} \right. \nonumber \\
 \mbox{} \left. + (v_{12})_{ij} (v^*_{22})_{kj}
 \right] = 0, \nonumber \\
 \sum_{j} \left[(u_{11})_{ij} (v_{21})_{kj} - (u_{12})_{ij} (v_{22})_{kj}
 - (v_{11})_{ij} (u_{21})_{kj} \right. \nonumber \\
 \mbox{} \left. + (v_{12})_{ij} (u_{22})_{kj}
 \right] = 0, \nonumber \\
\sum_{j} \left[(u_{21})_{ij} (u^*_{21})_{kj} - (u_{22})_{ij}
 (u^*_{22})_{kj} - (v_{21})_{ij} (v^*_{21})_{kj} \right. \nonumber \\
 \mbox{} \left. + (v_{22})_{ij} (v^*_{22})_{kj}
 \right] = -\delta_{ik}, \nonumber \\
\sum_{j} \left[(u_{21})_{ij} (v_{21})_{kj} - (u_{22})_{ij} (v_{22})_{kj}
 - (v_{21})_{ij} (u_{21})_{kj} \right. \nonumber \\
 \mbox{} \left. + (v_{22})_{ij} (u_{22})_{kj}
 \right] = 0.
 \nonumber \\
\label{B4}
\end{eqnarray}
Identities given in Eq. (\ref{B4}) are useful for controlling
precision of numerical integration of the corresponding
differential equations.


\end{document}